\documentclass[structabstract]{aa}   
\usepackage{graphicx}
\usepackage{txfonts}

\usepackage{natbib}
\bibpunct{(}{)}{;}{a}{}{,} 
\begin{document}
\authorrunning{M.W. Blanco et al.}
\titlerunning{Extended emission of four obscured post-AGB candidates}
   \title{VISIR-VLT high resolution study of the extended emission of
     four obscured post-AGB candidates\thanks{Based on
       observations collected at the European Organisation for Astronomical Research in the Southern Hemisphere, Chile. Program: 087.D-0367(A).}} 


\author{M.W. Blanco
          \inst{1},
           M.A. Guerrero
           \inst{1},
           G. Ramos-Larios
           \inst{2},
           L.F. Miranda
           \inst{3,4},
           E. Lagadec
          \inst{5},
          O. Su\'arez
          \inst{6}, 
          \and
          J.F. G\'omez
          \inst{1}   
          }

   \institute{Instituto de Astrof\'isica de Andaluc\'ia (IAA-CSIC),
              Glorieta de la Astronom\'ia S/N, 18008 Spain\\
              \email{mblanco@iaa.es, mar@iaa.es, jfg@iaa.es}
         \and 
              Instituto de Astronom\'\i a y Meteorolog\'\i a,
              Av.\ Vallarta No.\ 2602, Col.\ Arcos Vallarta, 44130 Guadalajara,
              Jalisco, Mexico\\
              \email{gerardo@astro.iam.udg.mx}
         \and
             Universidad de Vigo, Departamento de F\'isica aplicada,
             Facultad de Ciencias, Campus Lagoas-Marcosende s/n, 36310
             Vigo, Spain\\ 
             \email{lfm@iaa.es}
         \and
             Consejo Superior de Investigaciones Cient\'ificas (CSIC),
             Madrid, Spain
         \and 
              European Southern Observatory (ESO), 
              Karl Schwarschild Str. 2, Garching bei M\"unchen, Germany\\
              \email{elagadec@iaa.es}
         \and
            Laboratoire Lagrange, UMR7293, Universit\'e de Nice Sophia-Antipolis,
            CNRS, Observatoire de la C\^ote d'Azur, 06300 Nice, France\\
             \email{olga.suarez@unice.fr}
             }
   \date{Received September 15, 1996; accepted March 16, 1997}

 
  \abstract
{
The onset of the asymmetry of planetary nebulae (PNe) is expected to occur during
the late Asymptotic Giant Branch (AGB) and early post-AGB phases of low- and 
intermediate-mass stars.  
Among all post-AGB objects, the most heavily obscured ones
might have escaped the selection criteria of previous studies detecting 
extreme axysimmetric structures in young PNe.
}
{
Since the most heavily obscured post-AGB sources can be expected to
descend from the most massive PN progenitors, these should exhibit
clear asymmetric morphologies. 
High-resolution observations of these sources should reveal marked
bipolar morphologies, confirming the link between progenitor mass and
nebular morphology. 
}
{
We have obtained VISIR-VLT mid-IR images of a sample of four heavily obscured 
post-AGB objects barely resolved in previous \textit{Spitzer} IRAC observations 
in order to analyze their morphology and physical conditions across the mid-IR. 
The images obtained in four different mid-IR filters have been deconvolved, flux calibrated, and used to construct RGB composite pictures as well as color 
(temperature) and optical depth maps that allow us to study the
morphology and physical properties of the extended emission of the sources in our sample.
}
{
We have detected extended emission from the four objects in our sample
and resolved it into several structural components that
are greatly enhanced in the
temperature and optical depth maps. 
The morphologies of the sample, as well as their physical conditions, 
reveal the presence of asymmetry in three young PNe (IRAS\,15534$-$5422, 
IRAS\,17009$-$4154, and IRAS\,18454$+$0001), where the asymmetries can 
be associated with dusty torii and slightly bipolar outflows. 
The fourth source (IRAS\,18229$-$1127), a possible post-AGB star, is better described as a rhomboidal 
detached shell. 
}
{
The heavily obscured sources in our sample do not show
extreme axisymmetric morphologies. This is at odds with the expectation of highly asymmetrical
morphologies in post-AGB sources descending from massive PN
progenitors,
which is otherwise supported by observations of bright mid-IR unobscured 
sources.  
The sources presented in this paper may be sampling critical early phases 
in the evolution of massive PN progenitors, before extreme asymmetries 
develop.  
}


   \maketitle
%

\section{Introduction}

Evolved low- and intermediate-mass stars ($0.8 M_\odot < M < 8  M_\odot $) 
undergo  heavy mass-loss episodes during the red giant and
  Asymptotic Giant Branch (AGB) phases. 
At the tip of the AGB, these stars will eject most of their stellar
  envelopes in a short time until they evolve into 
  post-AGB stars, the immediate precursors of planetary nebulae
  (PNe). PNe display an impressive variety of
  morphologies, from spherical to the most complex shapes: bipolars and
  multipolars with point-symmetric structures and collimated jets moving at
  high velocities.
It has been suggested that asymmetric PNe are the rule, rather than
the exception \citep{manchado2000,schwarz93}.
The shaping mechanism of complex PNe is a matter of intense 
debate and is not yet completely understood \citep{balick2002}

\begin{figure*}
   \centering
   \includegraphics[height=12cm,bb=-30 250 680 718]{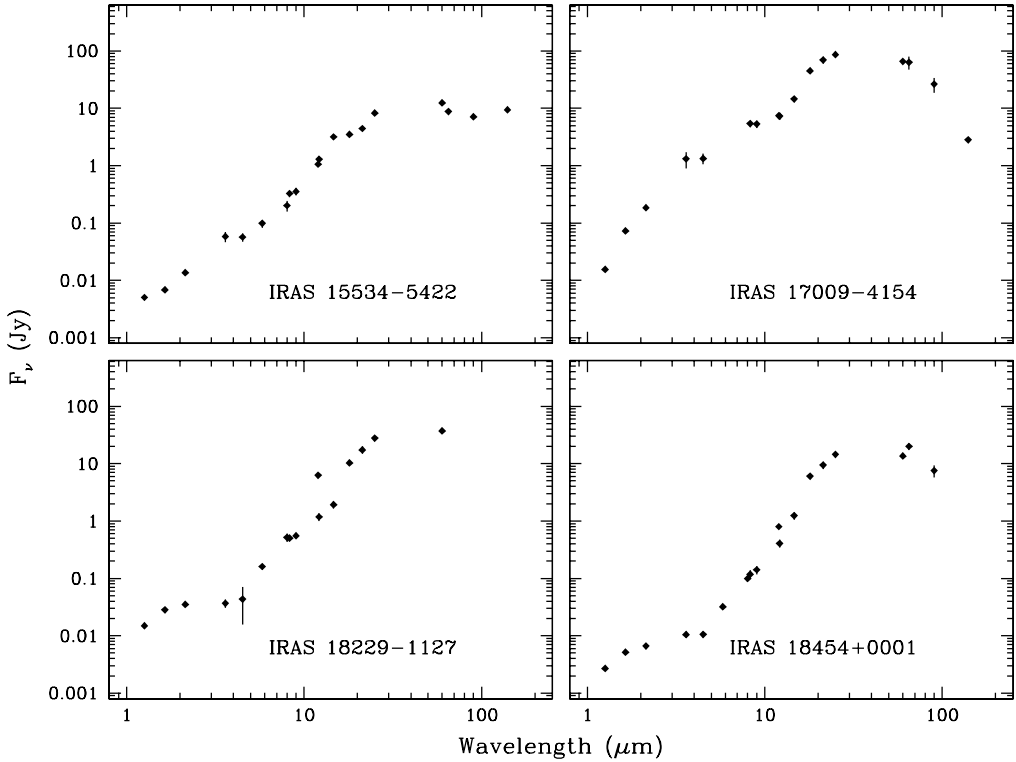}
   \caption{ Near- and mid-IR spectral energy distributions (SEDs) of the
     sources in our sample. See \citet{RL09,RL2012} for further
     details on the different datasets used to built these SEDs.}
              \label{Fig1}
    \end{figure*}

It has been claimed that the short transition between the AGB and
post-AGB phase contains the clues to understand the transformation of 
the spherical stellar envelope into an asymmetric PN \citep{Sahai98}.  
AGB stars are surrounded by thick and compact circumstellar envelopes 
rich in dust; this makes their detection almost impossible at optical
 wavelengths, whereas their infrared emission is strong.
As they evolve into the post-AGB phase, the temperature of the stellar 
cores increases and the envelopes become optically thin, although it 
must be noted that there are sources at this stage and even young PNe that are significantly
obscured because they preserve optically thick envelopes \citep[e.g., IRAS\,17347$-$3139,][]{Greg2004}.
Observational studies of the reflection, thermal dust and ionized
emission around proto-PNe and young PNe 
\citep{Sahai98,Sahai07,Ueta2000,Lagadec2011,Sahai2011} 
\emph{typically detect} the existence of extreme axisymmetric morphologies, with 
highly collimated bipolar lobes and equatorial rings or torii.  
On the other hand, spherical and elliptical morphologies seem to
be rare among the objects in this transition phase.

Infrared surveys (e.g., 2MASS, \emph{IRAS}, and \emph{AKARI}) have
contributed to the study of objects in the late AGB and early
post-AGB stages,  providing the first insights on this short
transition phase \citep[e.g.,][]{Garcia97,JEetal06,Coxetal11}.  
Recently, \citet[][hereafter RL09 and RL12, respectively]{RL09,RL2012}
have investigated a sample of heavily obscured post-AGB and PNe 
candidates selected according to their \emph{IRAS} colors and
the lack of an optical counterpart \citep{Suarez2006} using near-IR 
$JHK$ and \textit{Spitzer} IRAC GLIMPSE images, and \emph{MSX}, 
\emph{AKARI}, and \emph{IRAS} photometric data. 
RL09 and RL12 have gathered a small sample 
of four objects that are resolved in \textit{Spitzer} IRAC GLIMPSE images, 
namely IRAS\,15534$-$5422, IRAS\,17009$-$4154, 
IRAS\,18229$-$1127, and IRAS\,18454$+$0001.
The spatial resolution of these images ($\sim$2\arcsec), however, is
not adequate to investigate the angularly small structures typical of
sources in these evolutionary phases {\citep{Suarez2011,Lagadec2011}}. Mid-IR observations using the new generation of ground-based
telescopes provide a unique opportunity to resolve the extended 
emission detected in these four highly obscured post-AGB objects 
in their transition to the PN phase. 

In this paper we present mid-IR VISIR-VLT high angular resolution
observations of these four sources.  
The images have been used to describe their morphologies, to analyze their innermost structures, and to derive color (temperature) and optical depth maps.
We next describe the sample in Sect.~2, the observations and data 
reduction in Sect.~3, and present the results in Sect.~4. 
The results are discussed in Sect.~5 and a short conclusion is 
provided in Sect.~6.

\section{The sample}

RL09 and RL12 reported the detection of extended emission in
\emph{Spitzer} IRAC images of four post-AGB source candidates, namely
IRAS\,15534$-$5422, IRAS\,17009$-$4154, IRAS\,18229$-$1127, and 
IRAS\,18454$+$0001. 
According to the spectral energy distribution (SED) classification
scheme of post-AGB sources introduced by \citet{Veen89}, the SEDs of
these four sources shown in Figure~\ref{Fig1} can be assigned to Type
II for IRAS\,15534$-$5422 (peak at $\sim$25 $\mu$m and gradual 
fall-off to shorter wavelengths, although we note that its SED
also suggests a near-IR excess at 2--5 $\mu$m), to Type III for IRAS\,18229$-$1127 and
IRAS\,18454$+$0001 (peak at $\sim$25 $\mu$m and steep fall-off to 
shorter wavelengths with a plateau between 1 and 4 $\mu$m), and to 
Type IV for IRAS\,17009$-$4154 (with a main peak at 25 $\mu$m and a 
secondary peak blue-wards). 
These SEDs are suggestive of two dust components, cold dust in the 
thermal IR and hot dust obscuring the central star in the near-IR. 

Besides the information provided by RL09 and RL12, there is no
detailed study available in the literature for these sources. 
Based on their \emph{IRAS} colors, IRAS\,15534$-$5422 is classified 
as a PN candidate \citep{Preite88}, as recently confirmed by \citet{Parker2012} by means of spectroscopy, whereas IRAS\,18454$+$0001 is 
classified as a post-AGB star \citep{Garcia97}.
We note that the \emph{IRAS} selection criteria for these sources 
\citep{Suarez2006} may overlap with those of Young Stellar Objects 
(YSOs), although RL09 noted that these four objects are not located 
near star forming regions, neither they have been classified as YSOs 
in the literature. 
The lack of CO line emission  
in IRAS\,15534$-$5422 and IRAS\,17009$-$4154, and the detection of 
narrow ($\simeq$0.8 km~s$^{-1}$) CO line emission toward 
IRAS\,18454$+$0001 \citep[][RMS Survey]{Urqu2008} are inconsistent
with a YSO nature.  
Based on the \emph{IRAS} colors of the sample, we can also discard a 
possible symbiotic star nature, as these have typically values of the 
\emph{IRAS} [12]$-$[15] color $\sim$0.8 \citep{Kenyon1988}, whereas 
the sources in our sample present values $\geq$2.  
Moreover, our sources do not present variability in near-IR observations 
(RL09 and RL12), thus we can disregard an eruptive nature, as for 
instance in circumstellar shells around luminous blue variable sources 
\citep[e.g., IRAS\,18576$+$0341,][]{Buemi_etal10}.

The detection of radio continuum emission
in IRAS\,15534$-$5422, IRAS\,17009$-$4154, and IRAS\,18454$+$0001 \citep{Urqu2008}
confirms that ionization is already present, implying that these
three sources may have already entered the PN phase.
Indeed, Br$\gamma$ emission from ionized material has been 
detected in IRAS\,15534$-$5422 (RL12) and in IRAS\,17009$-$4154 
\citep{Steene2000}.  
For IRAS\,15534$-$5422, this emission is found to be extended and to
display a bipolar morphology (RL12).  
We are thus confident on the PN nature of IRAS\,15534$-$5422, 
IRAS\,17009$-$4154, and IRAS\,18454$+$0001.

As for IRAS\,18229$-$1127, the absence of filamentary diffuse emission in the mid-IR, which 
is characteristic of YSOs, and the similarities between its SED and 
that of IRAS\,18454$+$0001 can be used as additional arguments 
for a post-AGB classification. 


\section{Observations and data reduction}

\begin{table*}
\caption{Observations log}             
\label{table:1}      
\centering                          
\begin{tabular}{c c c c c c c c c c c c}        
\hline\hline                 
Object & $\alpha$ & $\delta$ & Date of &
\multicolumn{2}{c}{\underline{~~~~~Integration time~~~~}} & Standard star &\multicolumn{4}{c}{\underline{~~~~~~~~~~~~~~~~FWHM~~~~~~~~~~~~~~~~}} \\ 
& 
\multicolumn{2}{c}{(J2000.0)} & observation & N band & Q band & & PAH1 & SiC & [Ne\,{\sc ii}] & Q1 \\
& & & &  (s) & (s) & & (\arcsec)  & (\arcsec) & (\arcsec)  & (\arcsec) \\
\hline                        
   IRAS\,15534$-$5422 & 15$^{\rm h}$57$^{\rm m}$21$\fs$11 & --54$^{\circ}$30$^{\prime}$46$\farcs$4 & 2011-05-25 &  600 & 900 & HD\,133550 & 0.28 & 0.37 & 0.35 & 0.46 \\      
   IRAS\,17009$-$4154 & 17$^{\rm h}$04$^{\rm m}$29$\fs$6 & --41$^{\circ}$58$^{\prime}$38$\farcs$9 & 2010-05-24 &  300 & 450 & HD\,163376 & 0.31 & 0.36 & 0.38 & 0.48 \\
   IRAS\,18229$-$1127 & 18$^{\rm h}$25$^{\rm m}$45$\fs$0 & --11$^{\circ}$25$^{\prime}$56$\arcsec$ & 2010-06-14 &  600 & 600 & HD\,169916 & 0.34 & 0.51 & 0.42 & 0.49 \\
   IRAS\,18454$+$0001 & 18$^{\rm h}$48$^{\rm m}$01$\fs$5 & +00$^{\circ}$04$^{\prime}$47$\arcsec$ &2010-06-22 &  840 & 1320 & HD\,168723 & 0.28 & 0.35 & 0.35 & 0.49 \\
  
\hline                                   
\end{tabular}
\end{table*}

High angular resolution observations  (Program ID: 087.D-0367(A), PI: M.A. Guerrero) were obtained with the mid-IR
imager VISIR \citep{Lagage2004} attached to the Cassegrain focus of
Melipal (UT3) at the VLT.
The sources have been observed through four different
filters, PAH1
($\lambda_c$=8.54 $\mu$m, $\Delta\lambda$=0.42 $\mu$m), SiC
($\lambda_c$=11.85 $\mu$m, $\Delta\lambda$=2.34 $\mu$m) and
[Ne\,{\sc ii}] ($\lambda_c$=12.82 $\mu$m, $\Delta\lambda$=0.21 $\mu$m)
in the N band, and the Q1 filter
($\lambda_c$=17.65 $\mu$m, $\Delta\lambda$=0.83 $\mu$m) in the Q
band. The observations of standard stars were performed after every science
ObsBlock to correct for PSF artifacts and flux calibration purposes.
The date of observation, the integration time, and the FWHM of the
standard star observed subsequently are summarised in Table~\ref{table:1}.
The pixel scale is 0\farcs075 and the field of view (FoV) is 19\farcs2$\times$19\farcs2.

The data were taken using several exposures with
short DIT (Detector Integration Time) depending on the flux
of the source and the required S/N. The
observation mode used was the so-called NORMAL with a perpendicular
chop-throw of 8\arcsec. The chopping and nodding standard technique
was used to help in the removal of
the background signal. In the NORMAL mode used for these observations, all frames taken at
a chopping position are added immediately at the end of the exposure,
resulting in a data cube of reduced size.
The data reduction was carried out following standard
procedures of Gasgano-VISIR
pipeline (version 3.4.4), in which flat fielding correction, bad pixel removal, source
alignment, and
co-addition of frames are executed to produce a combined image
for each filter. The resulting FoV is $\leq$8\arcsec.

In order to improve the spatial resolution of the raw images, we
deconvolved each of them using its PSF observation according to two different
deconvolution algorithms: Maximum likelihood (number of iterations $\geq$5) and
Richardson-Lucy (number of iterations$\geq$ 10). 
The precise algorithm and number of iterations was determined by the
quality of the raw image, so that the spatial resolution of the
deconvolved image was
improved but no artifacts were introduced. The
maximum likelihood algorithm was used to deconvolve the
images of IRAS\,17009$-$4154, IRAS\,18229$-$1127 and
IRAS\,18454$+$0001, whereas the Richardson-Lucy algorithm was used for
IRAS\,15534$-$5422. The final images are presented in
Figure~\ref{Fig2}.

\begin{table*}
\caption{Estimated fluxes}             
\label{tableFlux}      
\centering                          
\begin{tabular}{l c c c c }        
\hline\hline                 
Object & PAH1 (8.54 $\mu$m) & SiC (11.9 $\mu$m) & [Ne\,{\sc ii}] (12.8
$\mu$m) & Q1 (17.7 $\mu$m) \\
& (mJy) & (mJy) & (mJy) & (mJy) \\
\hline                        
   IRAS 15534$-$5422 & 181.6$\pm$0.1 &  768.5$\pm$0.1  &
   1461.9$\pm$0.1 & 1891.9$\pm$0.2 \\      
   IRAS 17009$-$4154 & 9054.5$\pm$0.5 &  4212.2$\pm$0.1 & 3280.2$\pm$0.2 & 50674$\pm$5 \\
   IRAS 18229$-$1127 & 659.78$\pm$0.02 & 270.25$\pm$0.04 & 270.79$\pm$0.03 & 4101.8$\pm$0.1 \\
   IRAS 18454$+$0001 & 184.34$\pm$0.03 & 167.61$\pm$0.02 & 225.17$\pm$0.03 & 3096$\pm$1 \\
  
\hline                                   
\end{tabular}
\end{table*}

The final images have been flux calibrated using fluxes of the PSF
stars \citep{Cohen99}. For the flux calibration we have
performed aperture photometry (Table~\ref{tableFlux}).
We note that the angular extent of IRAS\,15534$-$5422 and
IRAS\,17009$-$4154 is similar to the final FoV of 8\arcsec. As the background
aperture may include emission from these objects, their fluxes listed
in Table~\ref{tableFlux} should be regarded as lower limits of the real fluxes.

\section{Results}
   \begin{figure*}
   \centering   
   \includegraphics[width=4.35cm,bb=14 14 355 355]{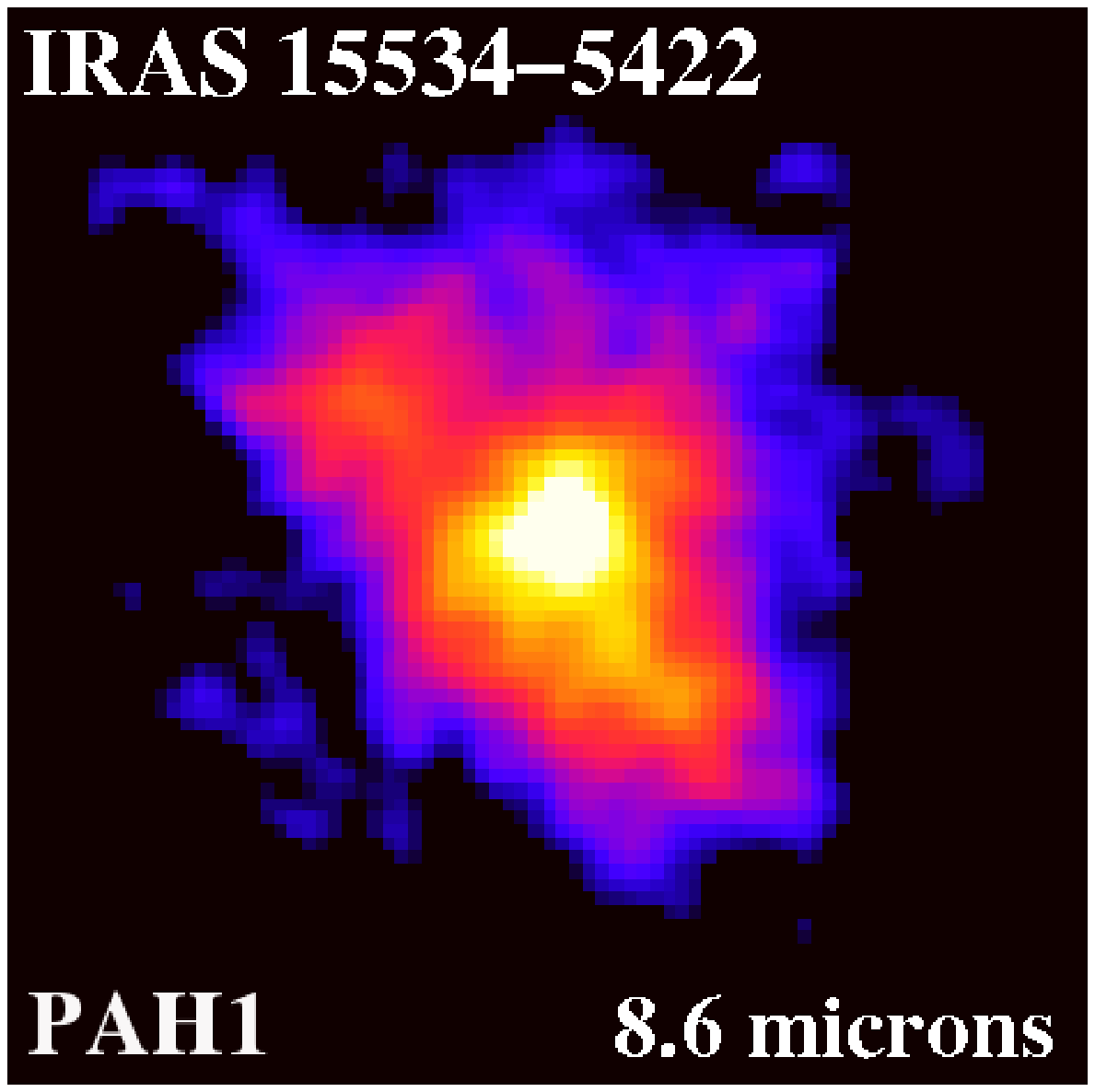}
   \includegraphics[width=4.45cm,bb=0 14 411 411]{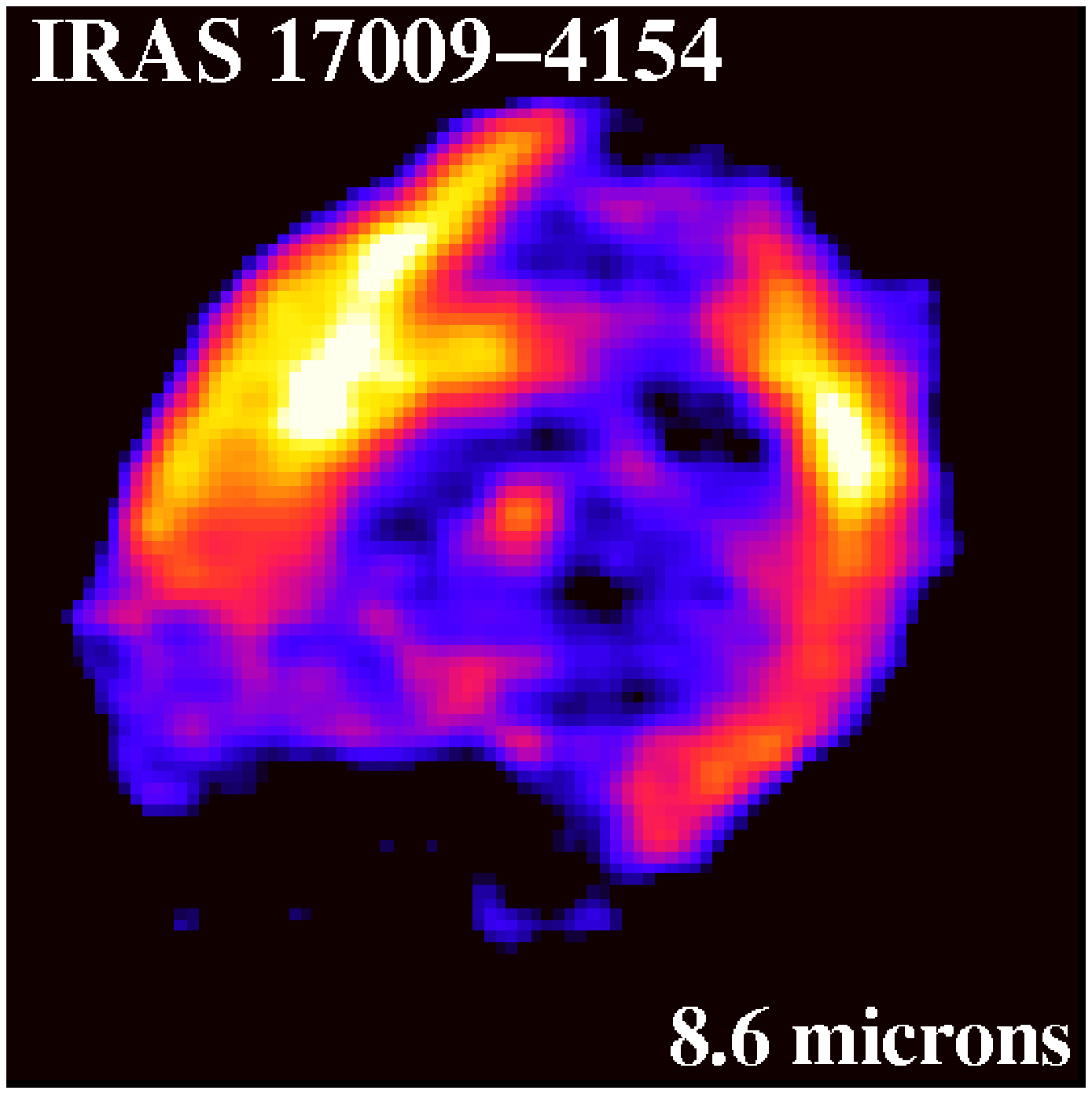}
   \includegraphics[width=4.4cm,bb=0 14 554 554]{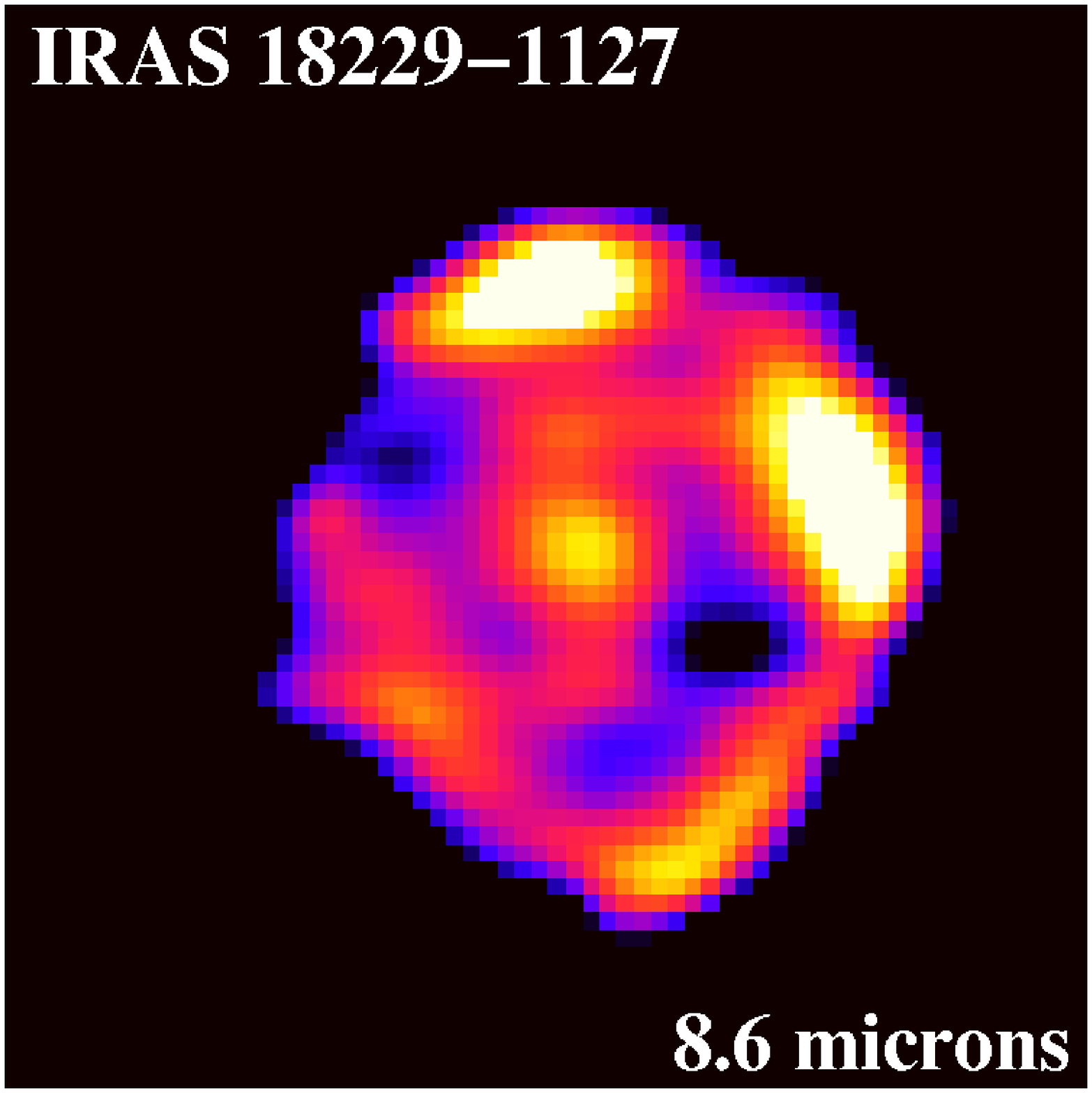}
   \includegraphics[width=4.45cm,bb=-14 0 530 530]{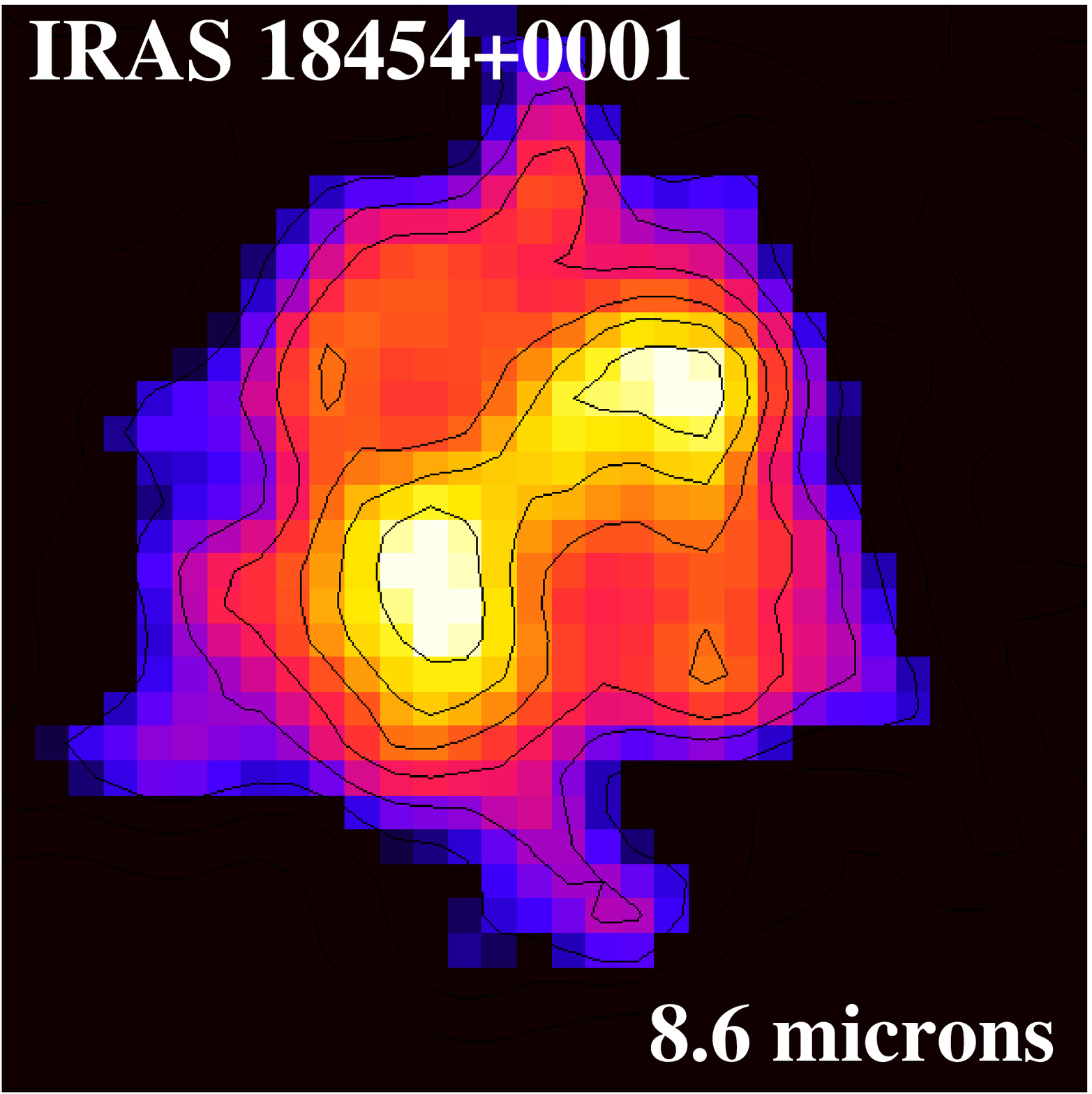}
   \includegraphics[width=4.35cm,bb=14 14 354 354]{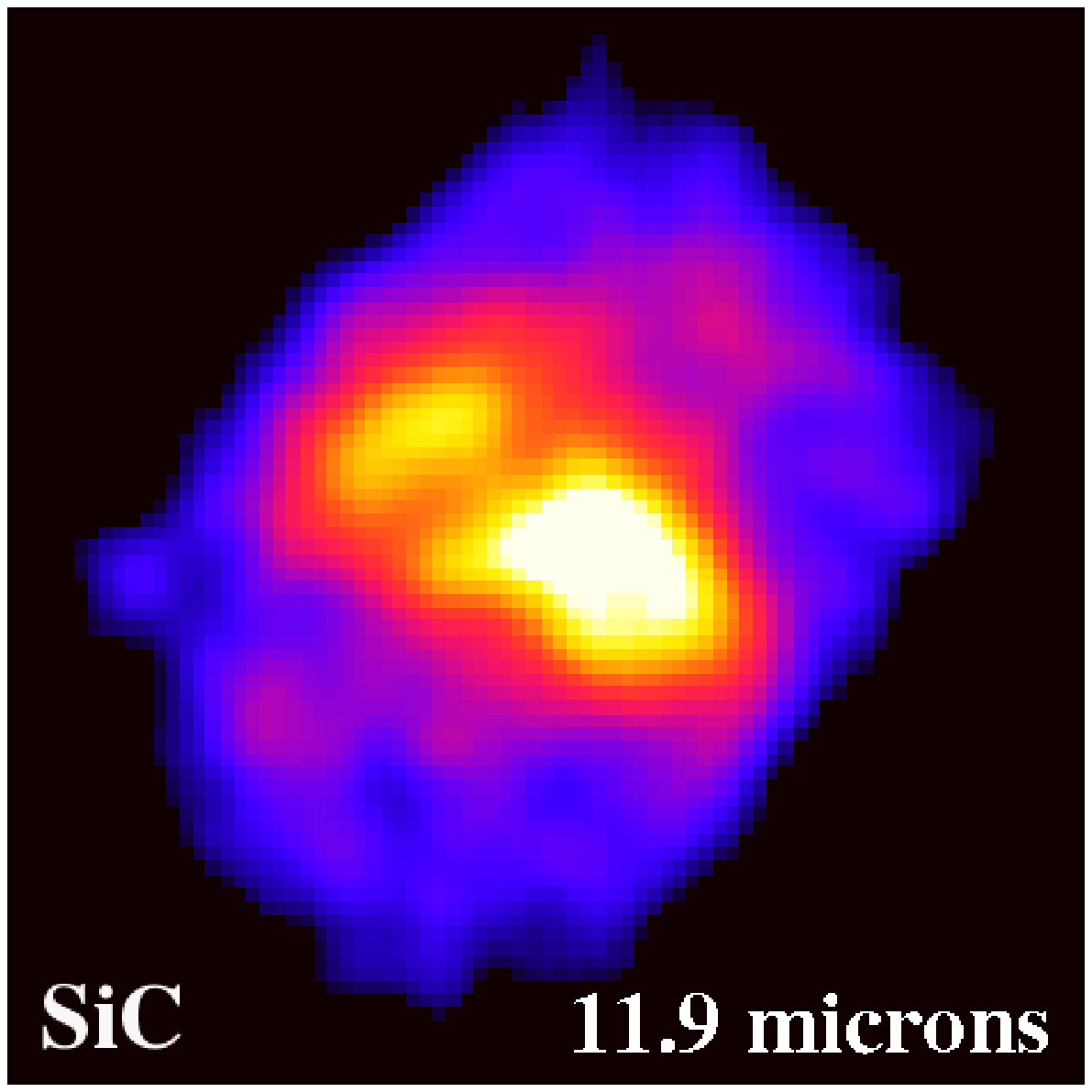}
   \includegraphics[width=4.45cm,bb=-14 0 394 394]{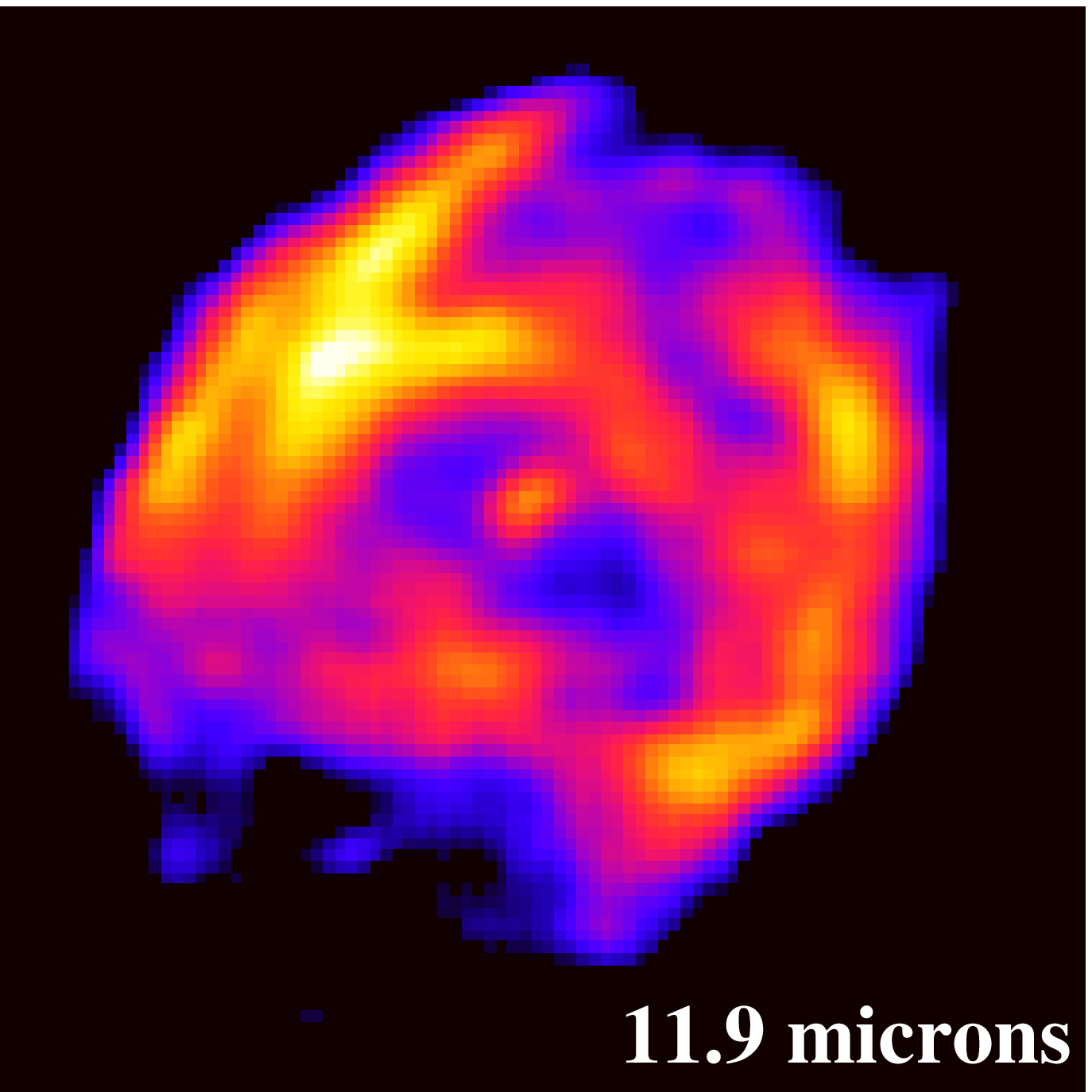}
   \includegraphics[width=4.4cm,bb=-14 0 530 530]{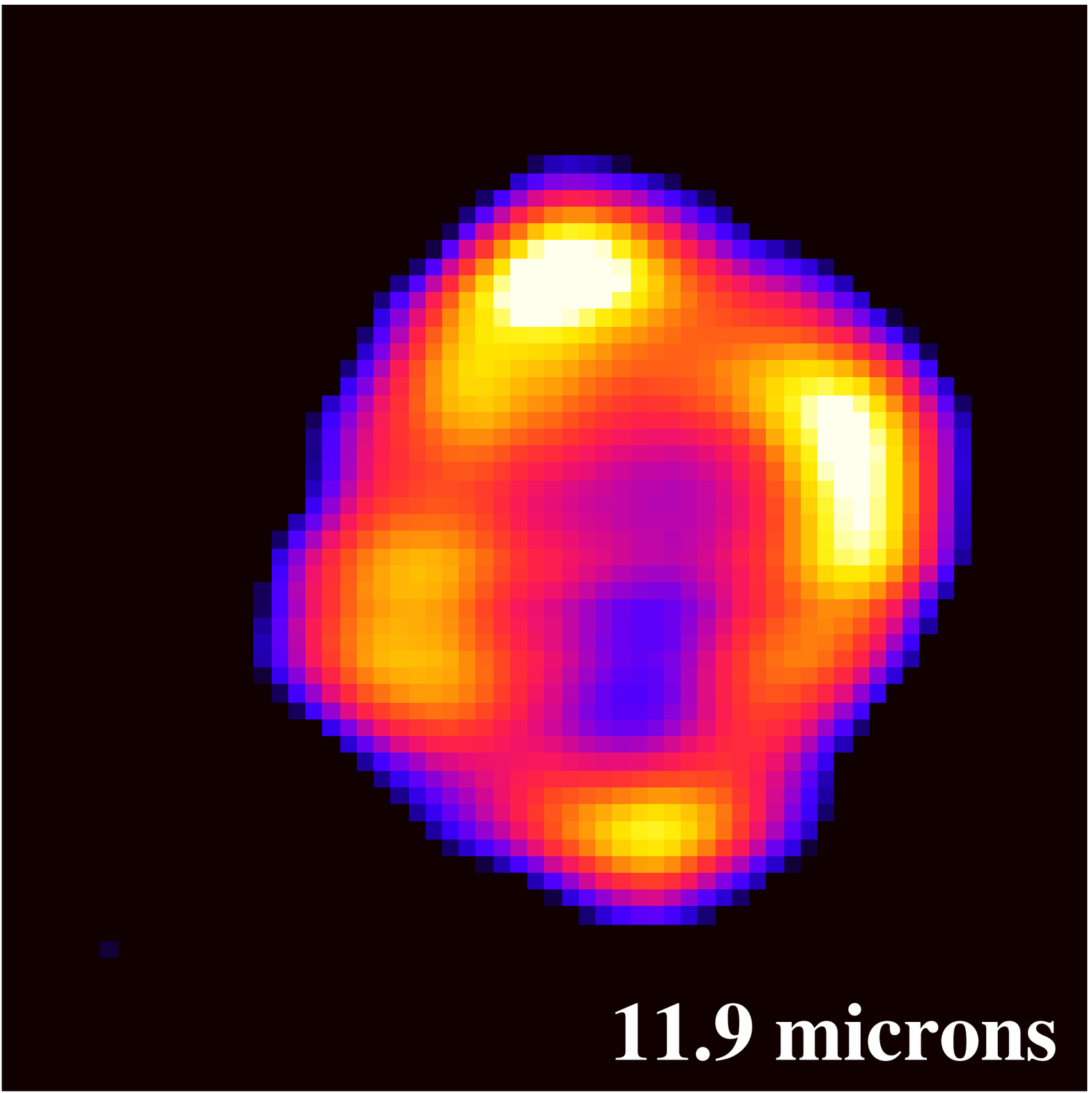}
   \includegraphics[width=4.43cm,bb=-14 0 530 530]{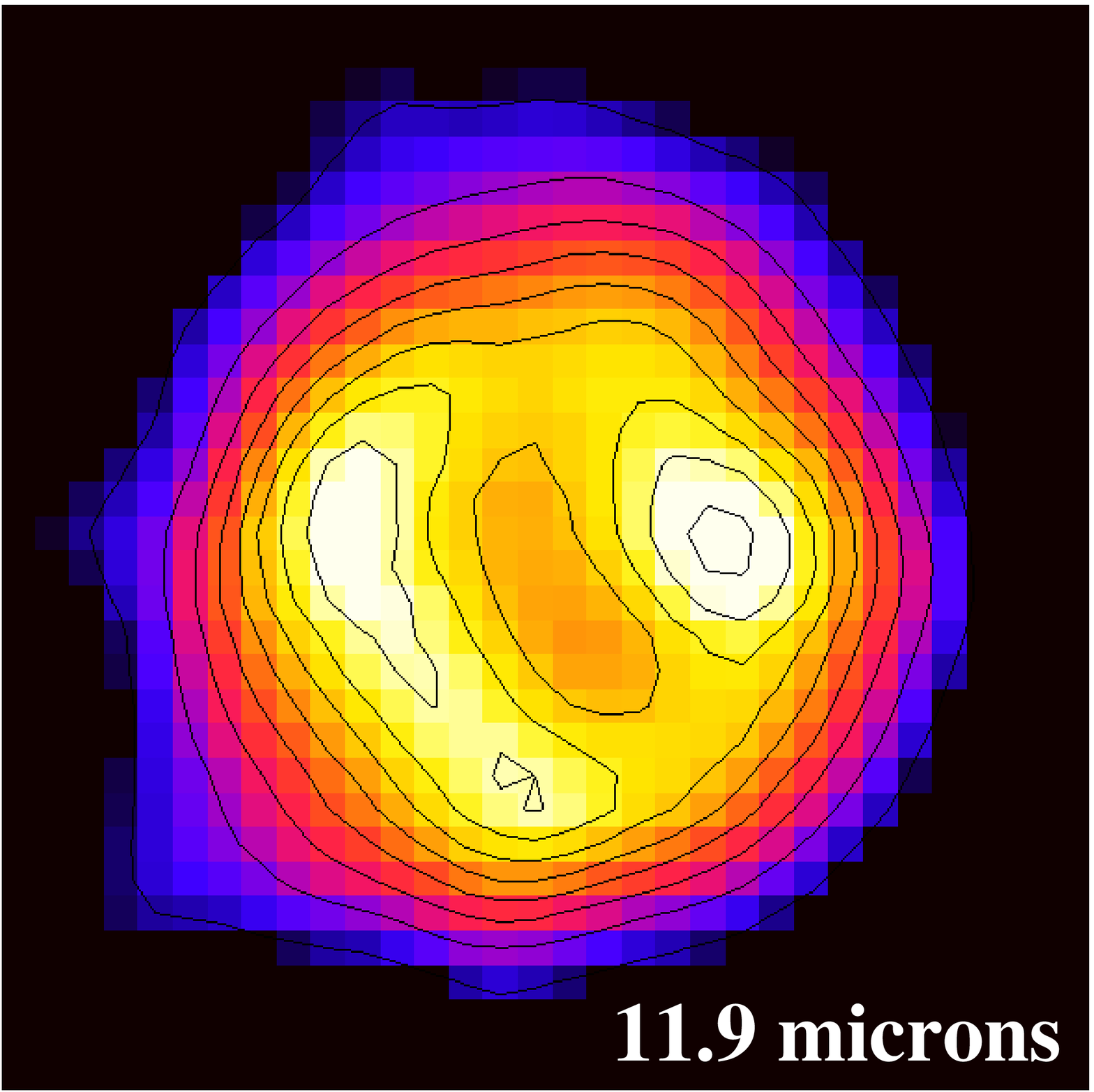}
   \includegraphics[width=4.3cm,bb=14 14 354 354]{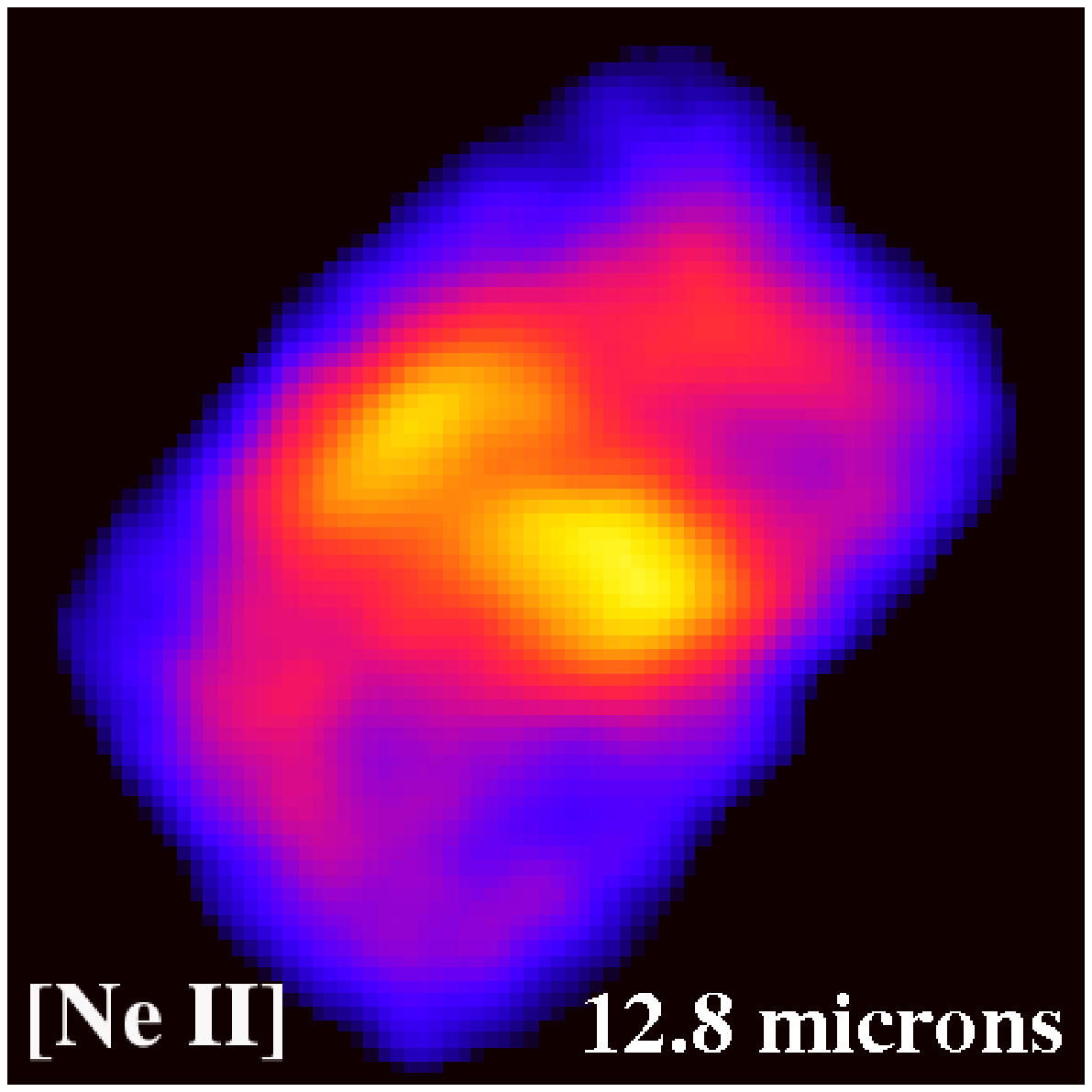}
   \includegraphics[width=4.42cm,bb=0 14 434 434]{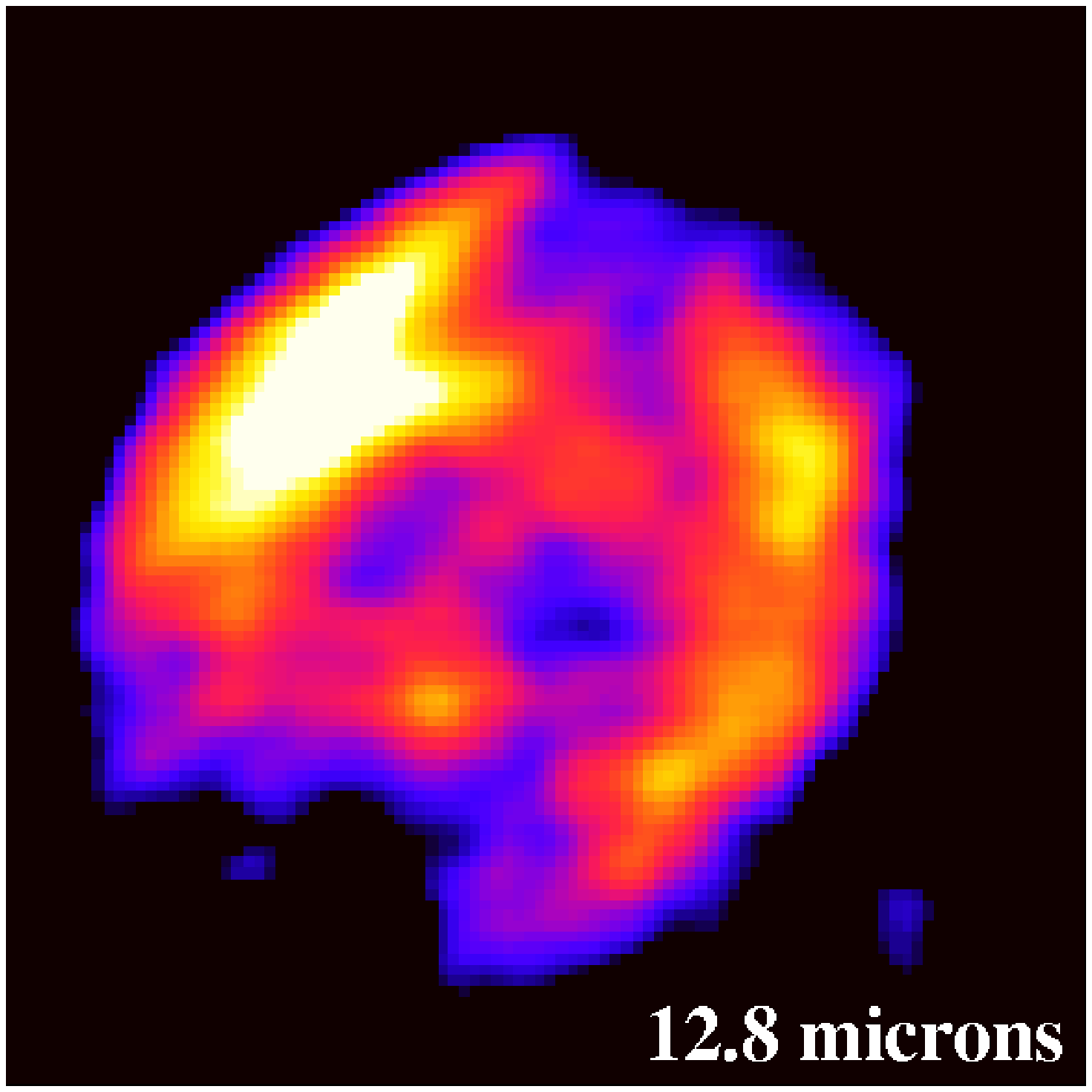}
   \includegraphics[width=4.4cm,bb=-1.5 14 553 553]{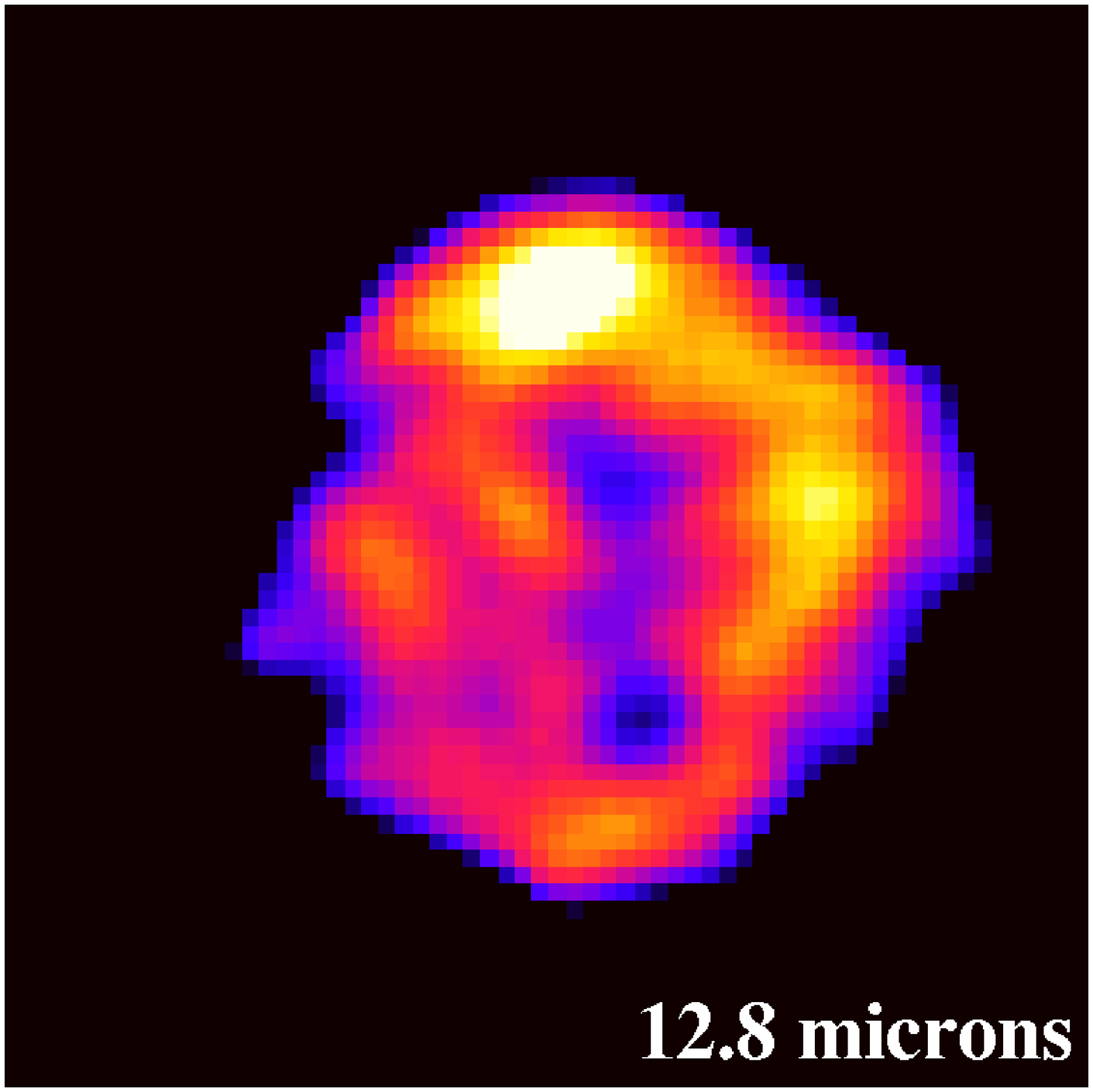}
   \includegraphics[width=4.41cm,bb=-14 0 531 531]{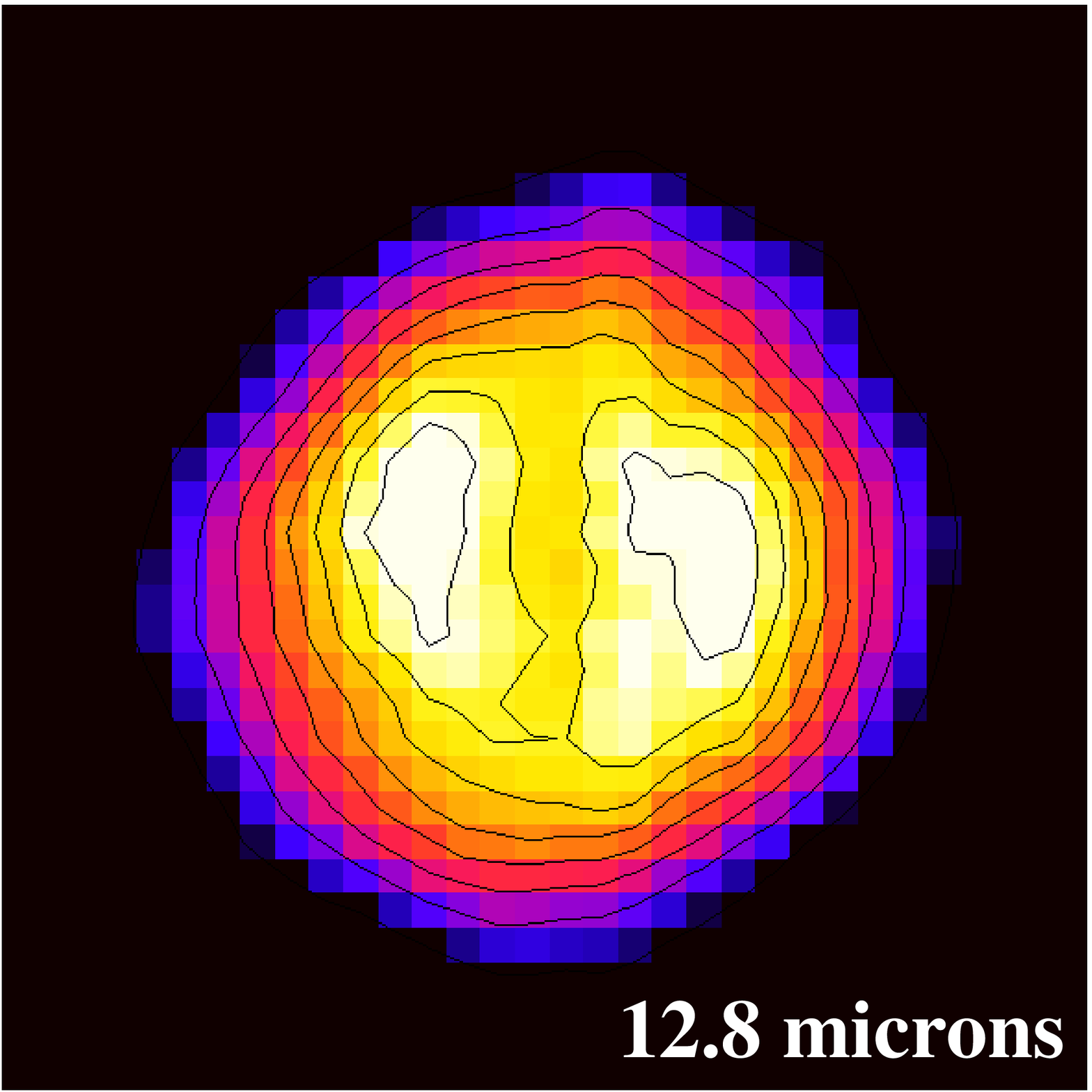}
   \includegraphics[width=4.3cm,bb=14 14 351 351]{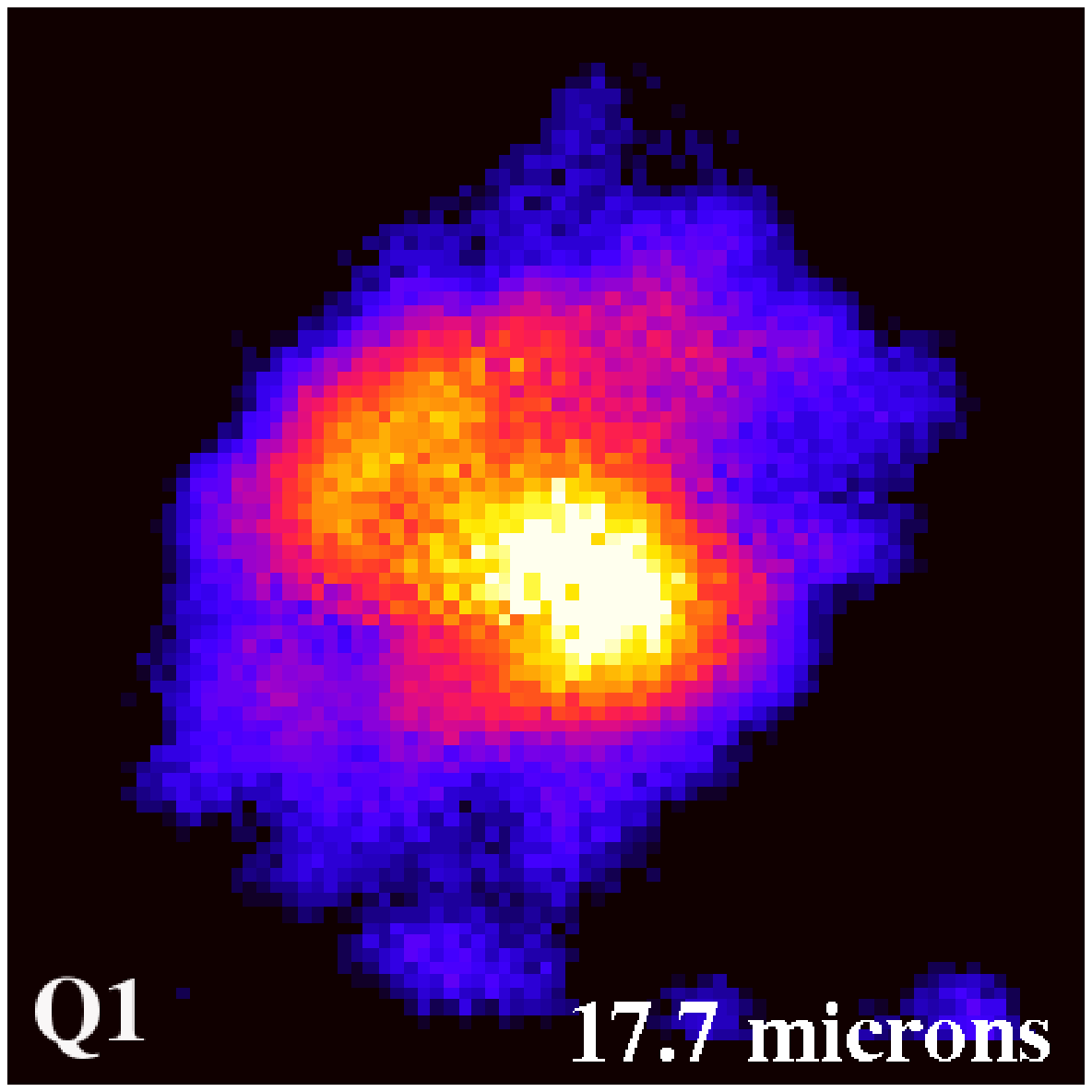}
   \includegraphics[width=4.45cm,bb=0 14 409 409]{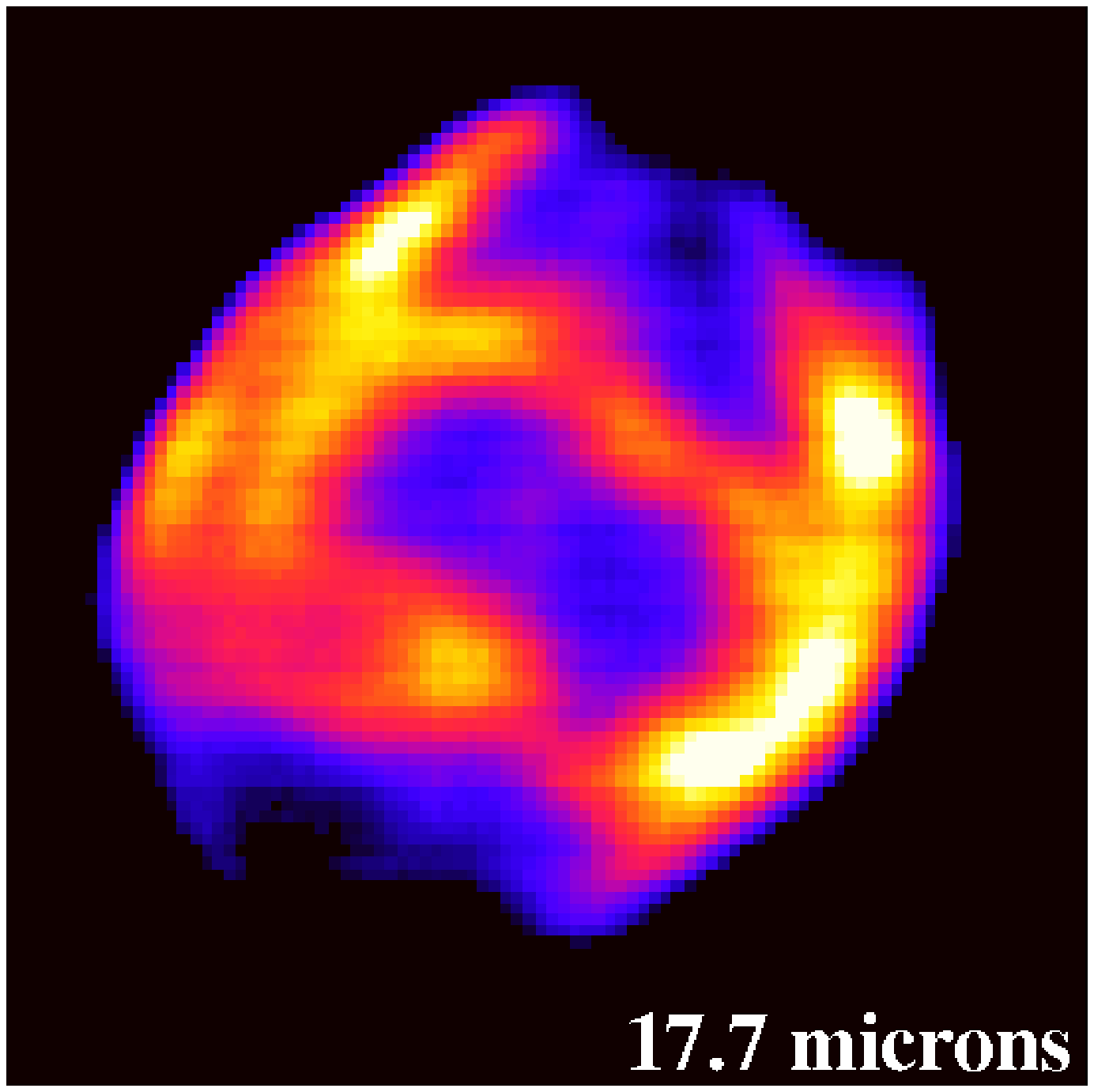}
   \includegraphics[width=4.41cm,bb=0 14 542 542]{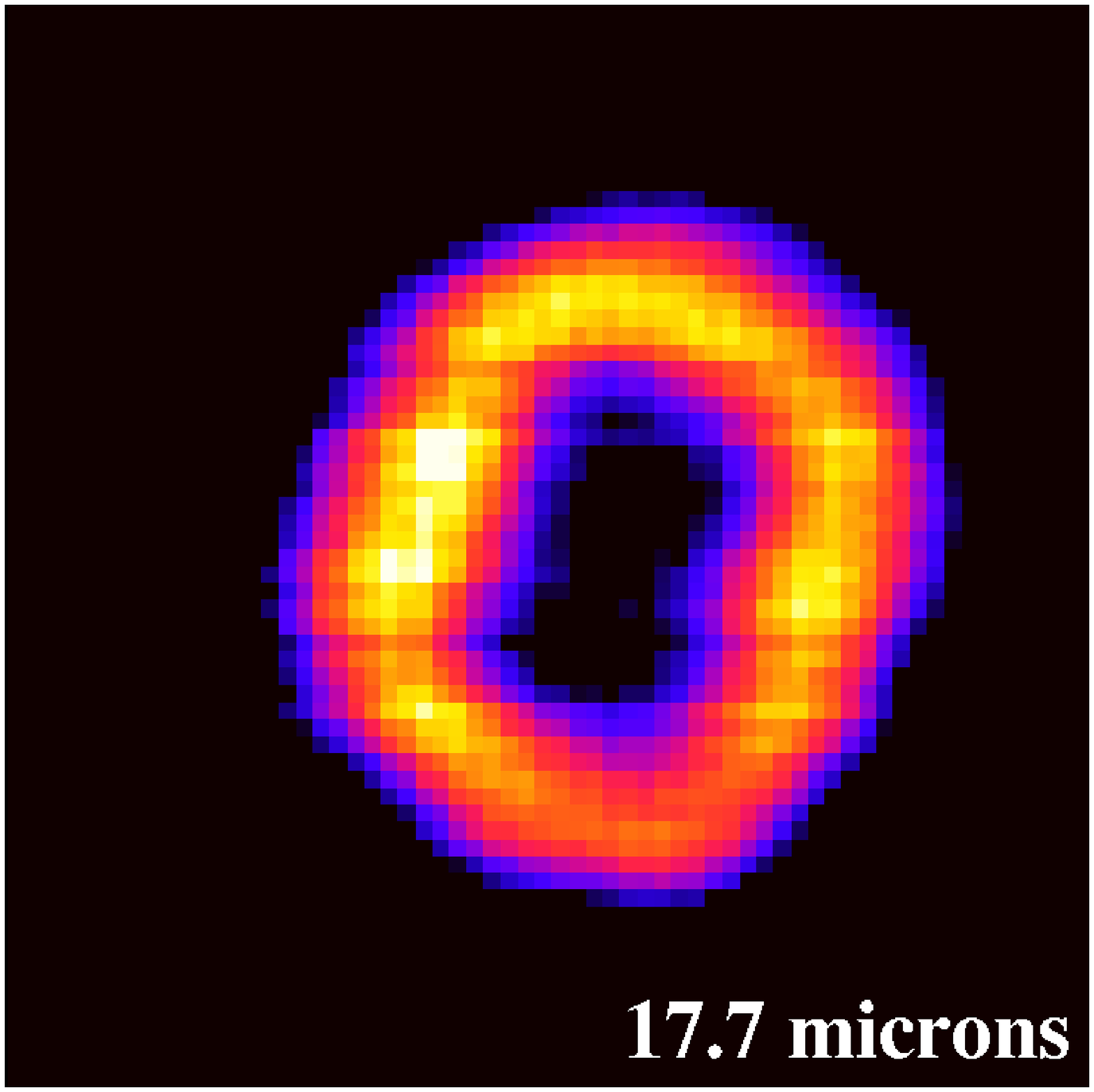}
   \includegraphics[width=4.42cm,bb=0 14 542 542]{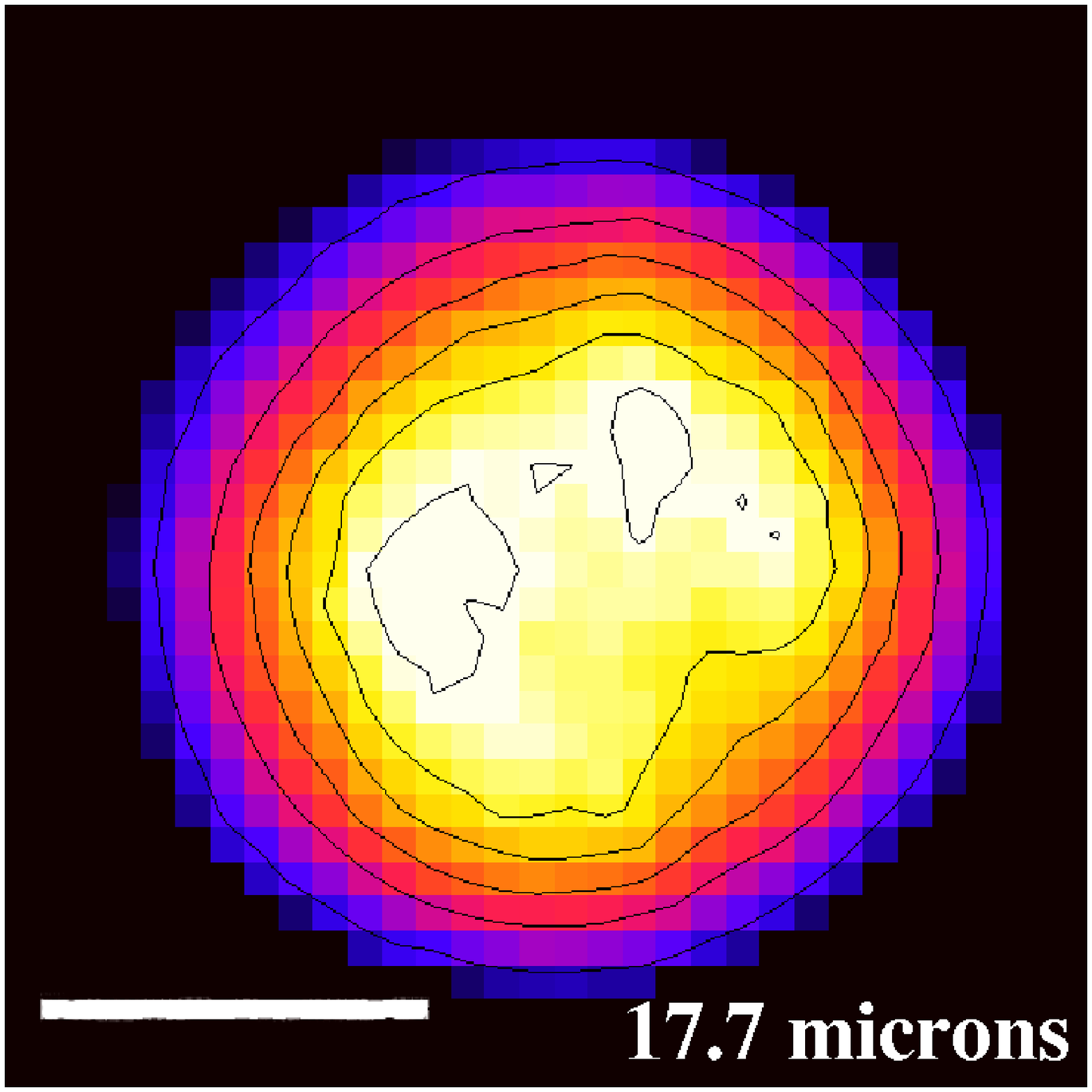}
   \includegraphics[width=4.3cm,height=4.08cm,bb=14 14 301 301]{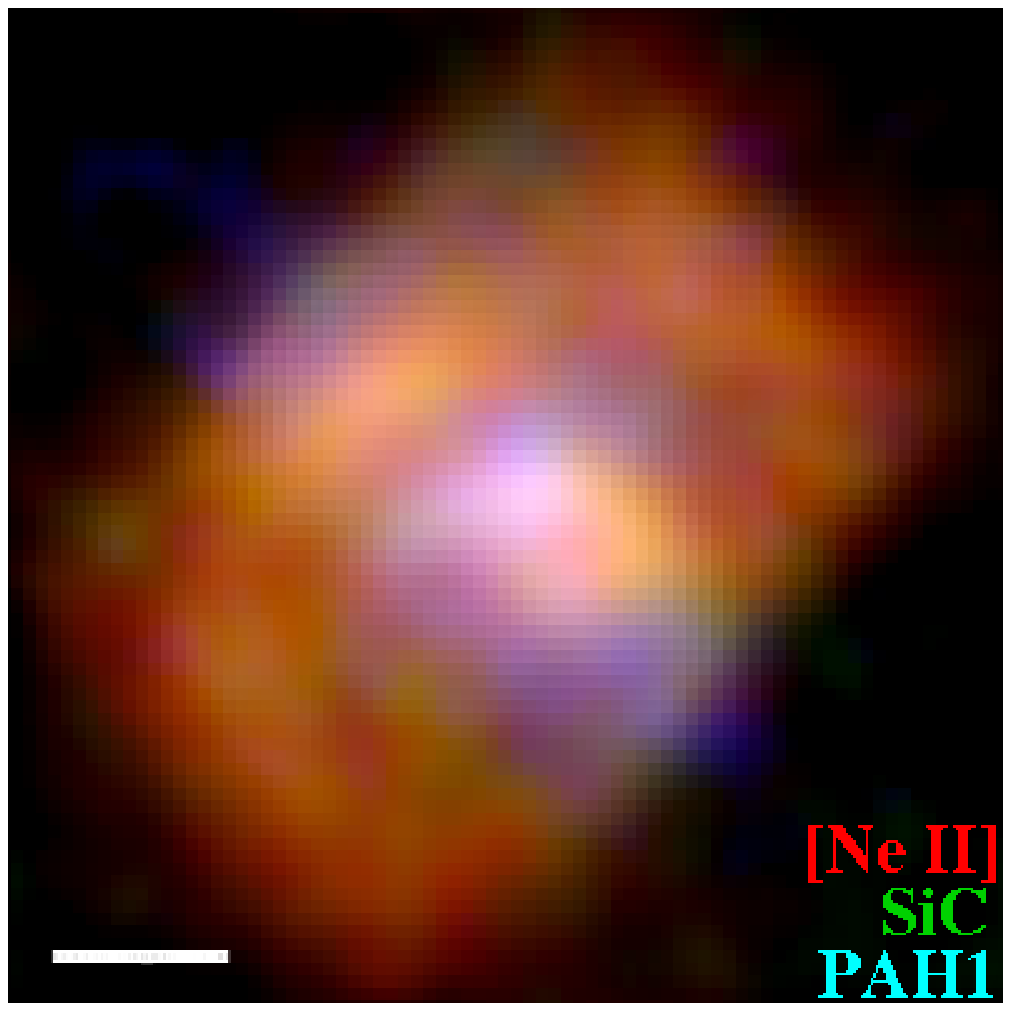} 
   \includegraphics[width=4.45cm,height=4.12cm,bb=0 14 349 349]{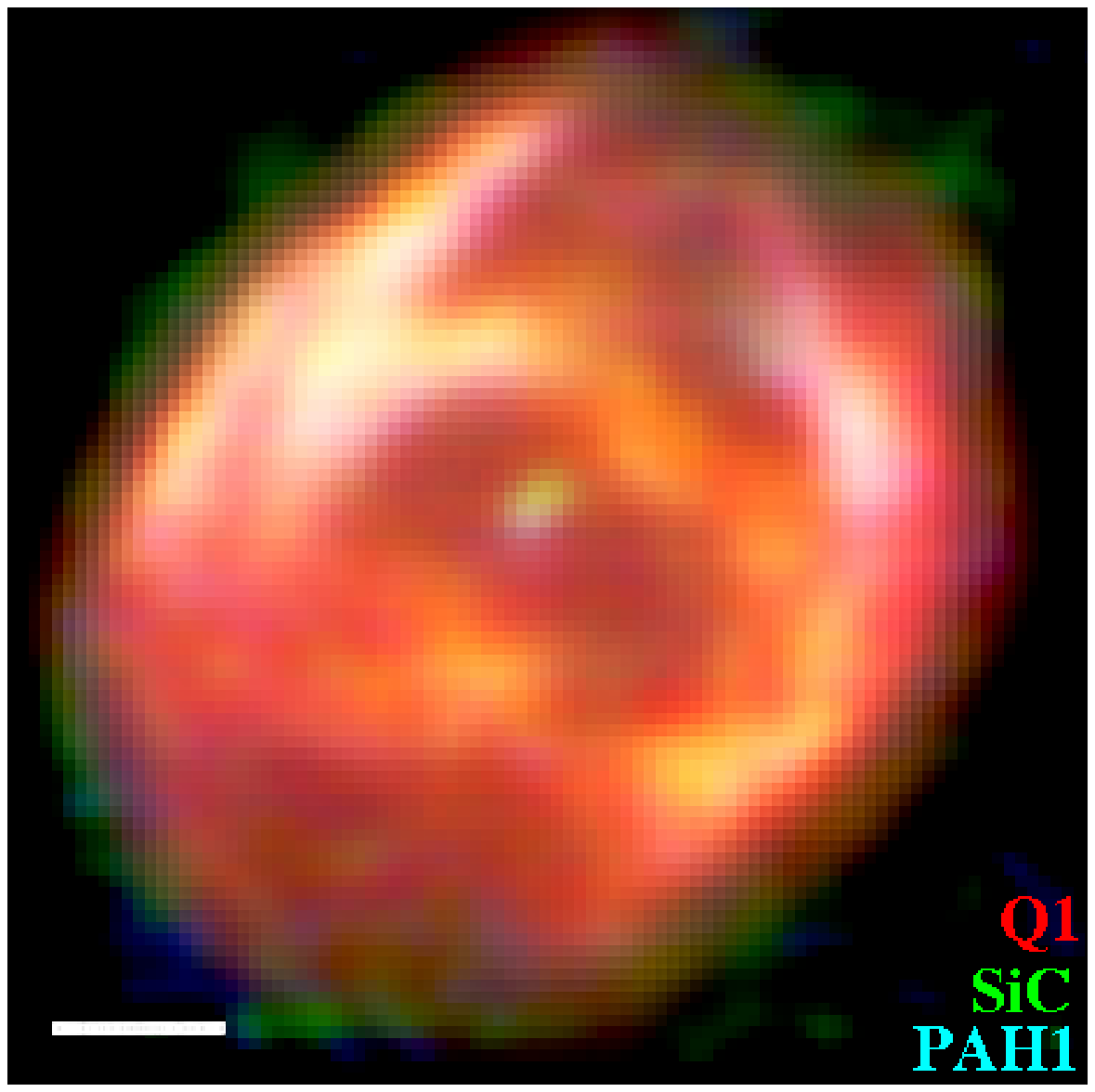}
   \includegraphics[width=4.48cm,height=4.145cm,bb=8 14 167
   167]{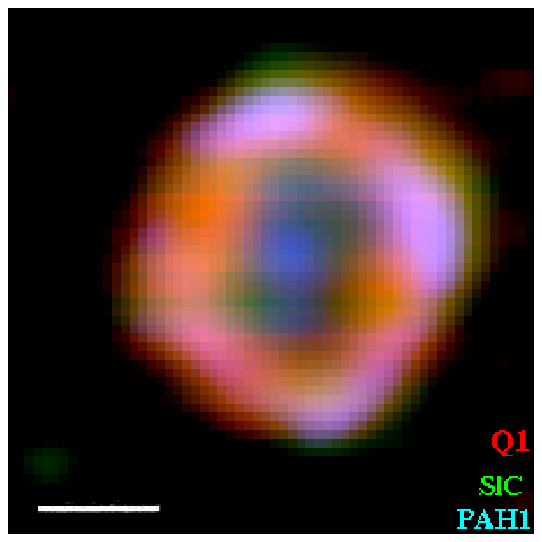}   
   \includegraphics[width=4.44cm,height=4.10cm,bb=12 14 166
   166]{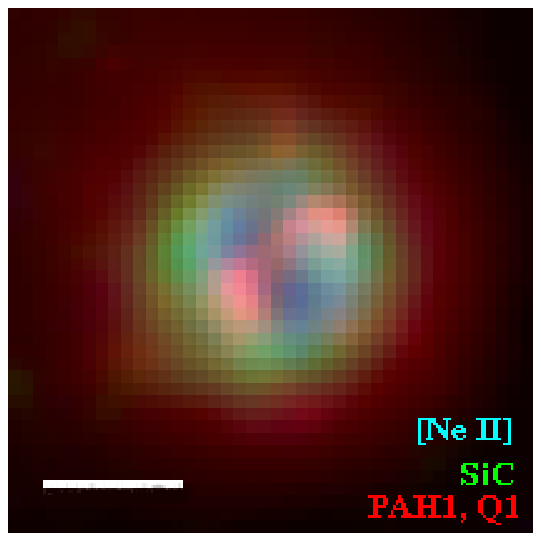} 
   \caption{Deconvolved VISIR images of the objects in our sample
     in the different filters observed in the N and Q bands. North is up and east to the
     left. The images of IRAS\,18454$+$0001 are overlaid with contours at 30, 40, 50, 60,
     70, 80, 90 and 100\% of the peak intensity to highlight its
     morphological features.The frames at the bottom
     are color-composite RGB pictures using the
     filters with the colors of the corresponding labels. The bar at
     the bottom of each RGB image and at the bottom of the Q1 image of
     IRAS\,18454$+$0001 represents 1$\arcsec$. 
The scale of the color-composite picture of IRAS\,18454$+$0001 is different from
that of the individual images to show in the former the faint halo
detected in the Q1 band.}

         \label{Fig2}
   \end{figure*}

%
The VISIR-VLT images presented in Figure~\ref{Fig2} reveal with
unprecedented detail the extended emission from these four sources.
These images also disclose notable variations among the morphological 
features shown in images obtained through different filters which we 
interpret to be related to the spectral features
registered by these filters. The emission in the PAH1 filter may
include the PAH1 feature at 8.6~$\mu$m associated with C-rich dust, as
well as thermal continuum emission. The broad SiC filter
includes thermal dust emission, silicate features, the PAH2 feature
at 11.3~$\mu$m, and emission lines
such as [Ne\,{\sc ii}] and [S\,{\sc iv}]. The [Ne\,{\sc ii}] filter is
fine tuned to the [Ne\,{\sc ii}] emission from ionized gas, but it may
also include the contribution from dust emission. Finally, the Q1
filter maps the continuum emission of the thermal dust at
17.7~$\mu$m.

\begin{figure*}
   \centering
   \includegraphics[height=7cm]{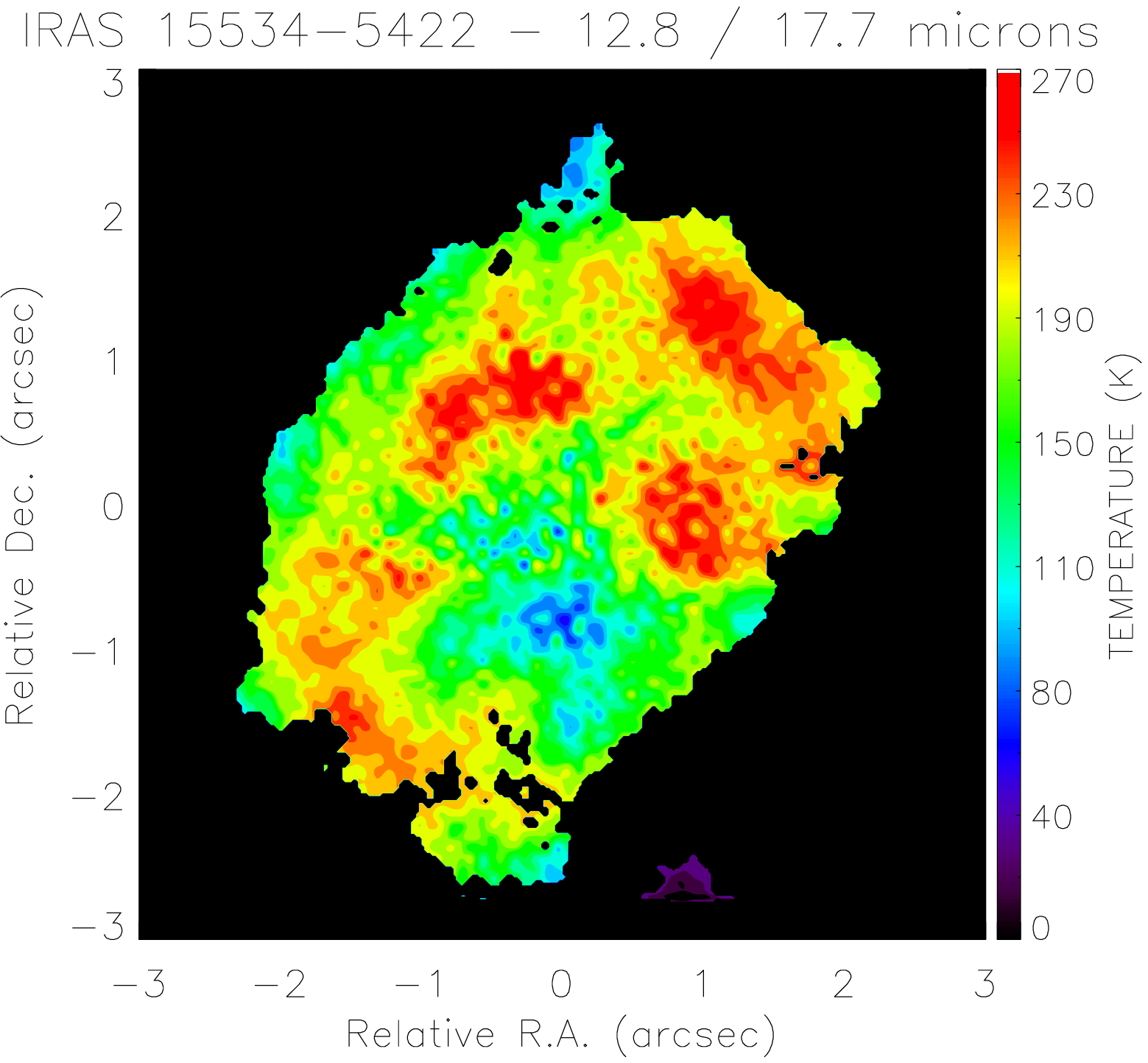}
   \includegraphics[height=7cm]{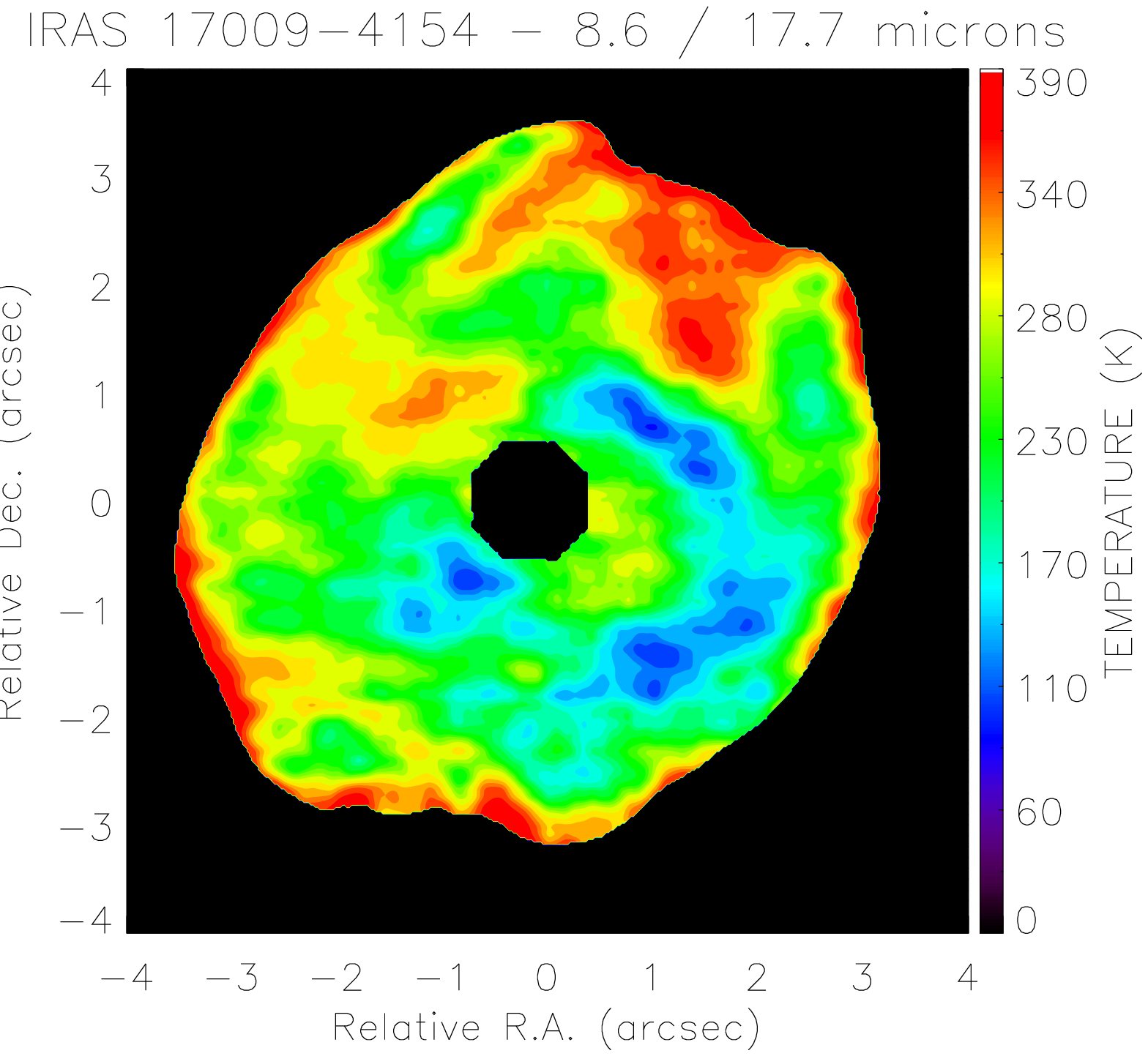}
   \includegraphics[height=7cm]{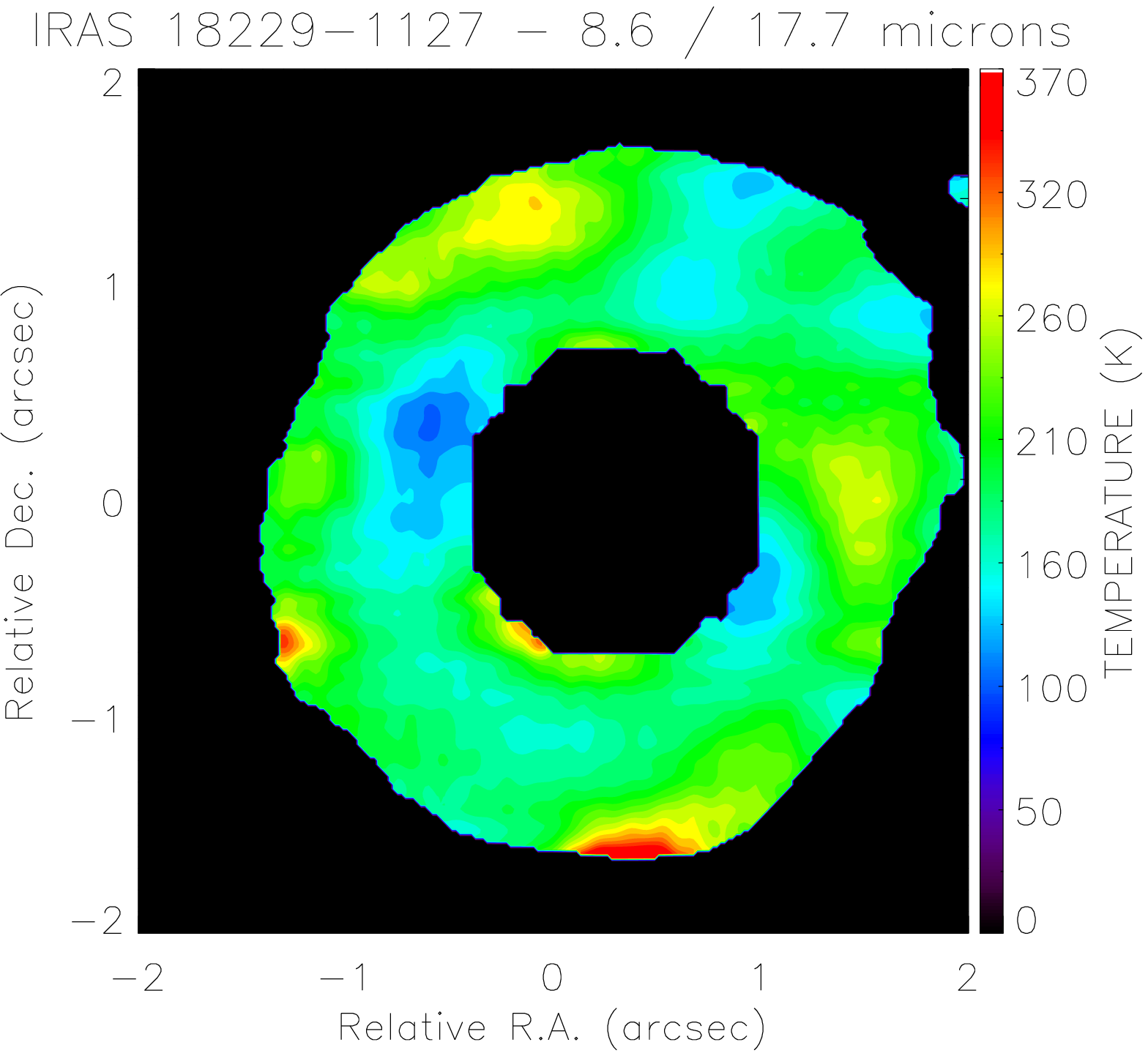}
   \includegraphics[height=7cm]{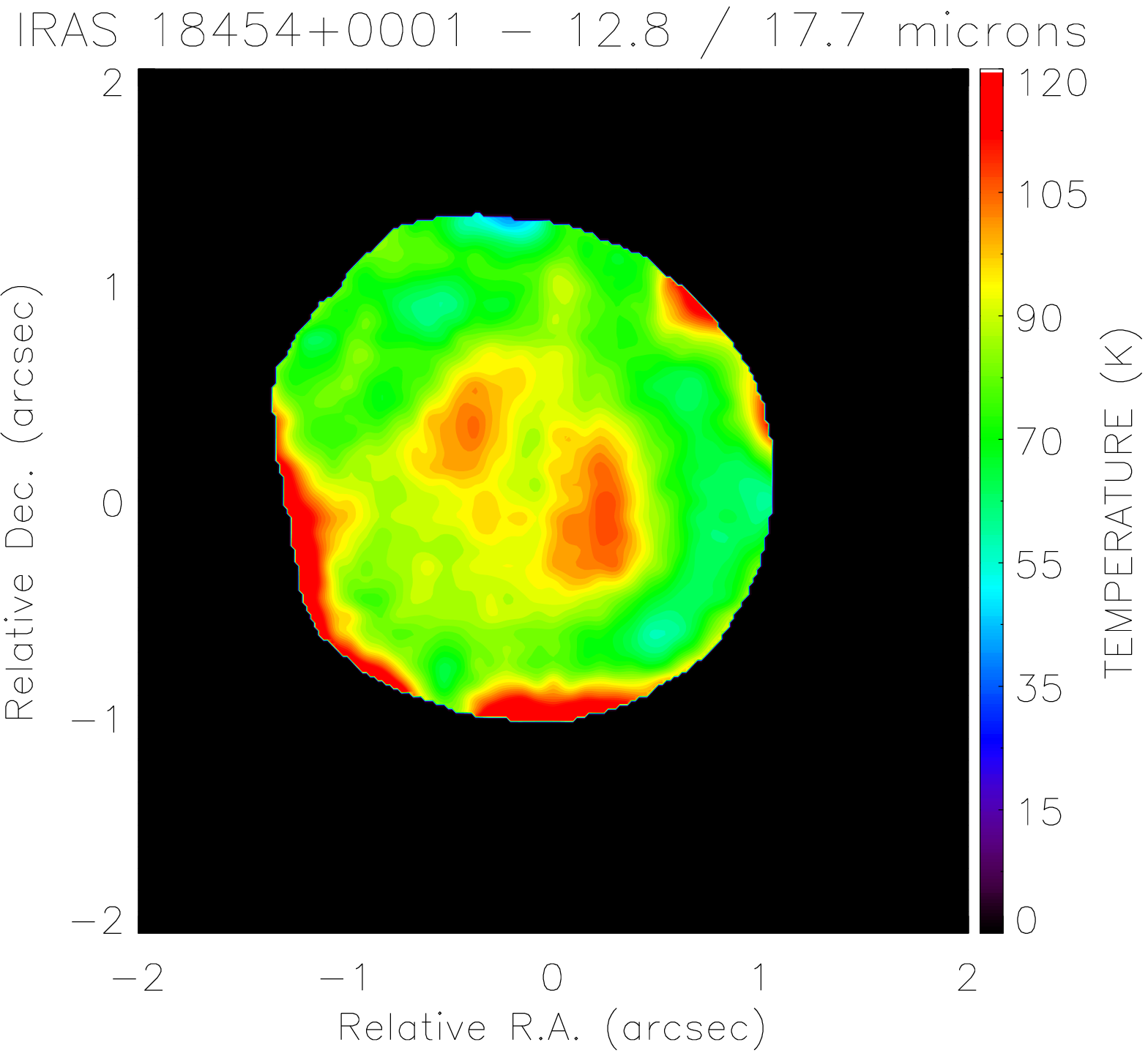}
      \caption{
Temperature (color) maps derived using the equation 1 of IRAS\,15534$-$5422, IRAS\,17009$-$4154,
IRAS\,18229$-$1127 and IRAS\,18454$-$18454.  
We note that the red patches at the borders of the color the maps of IRAS\,17009$-$4154 and
IRAS\,18454$+$0001 are artifacts caused by the reduced S/N 
ratio in the outermost regions of these sources. The central stars of
IRAS\,17009$-$4154 and IRAS\,18229$-$1127 have been masked to construct
the color maps
              }
         \label{Fig3}
   \end{figure*}

Considering that the thermal emission of the dust represents the
major contribution to the emission in the mid-IR we have applied
the procedure described by \citet{Dayal98}, we have
generated color maps (also known as temperature maps) using pairs
of flux calibrated images in different wavelengths for the four
objects in our sample (Figure~\ref{Fig3}).
This procedure relies on the relation between the intensity of the 
thermal dust continuum at a given wavelength ($I_\lambda$) with the
temperature and optical depth under the assumption of the optically 
thin emission of the warm dust component. 
As the extinction produced by cool dust component along the
line of sight can be considered to be similar at the mid-IR
wavelengths of the different filters used in this work, we can further
assume that its effects on the ratio maps are negligible. 
Since variations in the temperature produced by the emission
properties of the dust appear as color variations in the maps, the 
temperature of the dust can be approximated by the following expression:
\begin{equation}
T \approx \frac{1.44\times10^{4}(1/\lambda_2 -
  1/\lambda_1)}{ln[(I_{\lambda_1}/I_{\lambda_2})({\lambda_1} /
  {\lambda_2})^{3}]} {\mathrm K}
\end{equation}
where ${\lambda}_{1}$ and ${\lambda}_{2}$ are the two wavelengths
used for the estimation of the temperature, and $I_{\lambda_1}$ and $I_{\lambda_2}$ are the measured intensities at
these two wavelengths. This method assumes that the variation of the
emissivity ($Q$) can be
represented by a power law ($Q \sim {\lambda}^{-n}$)
and it has been proven to provide a good estimate of the 
spatial distribution of the cold and warm dust across the source 
studied \citep{Dayal98, Meixner99, Ueta2001, Lagadec2005}, being
helpful to reveal or enhance morphological features. 

In order to derive the color maps shown in Figure~\ref{Fig3},
we have used the Q1 image in conjunction with one N-band image, as the
longer wavelength range provides a better leverage of the continuum
slope which is determined by the thermal dust temperature. 
We have also avoided images in those filters that may present
important contributions of spectral features others than the continuum.  
Therefore, the SiC images have not been used, as they may
include multiple contributions from emission lines and dust features.

Once the value of the temperature is estimated, we can construct optical 
depth maps using the Planck function under the same assumption. 
Therefore, the optical depth at a certain wavelength $\lambda$ can be derived 
using the following expression:
\begin{equation}
\tau_\lambda \approx -ln \left[1-\frac{I_{\lambda}}{B_{\lambda}(T)}\right]
\end{equation}
where $I_{\lambda}$ is the intensity at the wavelength
analyzed. 
Using these maps, we are able to estimate the column density variations
across the nebula to assess whether the emission is optically thin at
the observed wavelength range. 

We next describe in detail the morphology, temperature and optical depth
properties of the individual sources.

\subsection{IRAS\,15534$-$5422} 

IRAS\,15534$-$5422 is resolved in at least two structures that are
distinguished by their morphology and emission properties
(Figure~\ref{Fig2}). 
A first structure is an elongated, bar-like feature with a size of 
$\sim$3\arcsec\ oriented at PA 55\degr. 
The emission of this bar is detected in PAH, SiC, and Q1, and 
is much weaker in [Ne\,{\sc ii}]. 
A second structure is an arc-like feature, $\sim$5\arcsec $\times$ 3\arcsec\
in size, which is oriented perpendicular to the bar and is particularly
prominent in [Ne\,{\sc ii}], but weak in SiC and Q1 and absent in PAH1.
This arc-like extended structure is not closed and presents a rectangular 
shape at low intensity levels.
The color composite picture of IRAS\,15534$-$5422 (Figure~\ref{Fig2}) 
suggests that the arc-like feature traces ionized material, while the bar 
is mostly dominated by thermal dust emission.

The color map of this source (Figure~\ref{Fig3}), obtained using the [Ne\,{\sc ii}] and the Q1
images, reveals a range of temperatures across the nebula from 80$\pm$1~K to 
270$\pm$1~K, with a mean value of $\approx$190~K. 
The color map greatly enhances the
arc-like feature, which is the hottest component, whereas the bar, with lower
temperatures, almost disappears in this map.
The optical depth maps at different wavelengths
(Figure~\ref{Fig4}), computed for a mean temperature value of 190~K, imply that the south-west
tip of the bar represents the highest density zone in this object. 
The [Ne\,{\sc ii}] optical depth map reveals a noticeable enhancement
of the column density at the north-east tip of the bar, where it goes
across the arc-like feature.

\subsection{IRAS\,17009$-$4154}

IRAS\,17009$-$4154 has an elongated morphology toward the north-west
with a size of 6$\farcs$3$\times$5$\farcs$5
in size. There is an innermost equatorial enhancement that can be appreciated
as an elongated ring-like structure ($\sim$3\arcsec$\times$2\arcsec) at PA$\sim$64\degr\ surrounding a central
star which is detected in all VISIR bands (Figure~\ref{Fig2}). 
Two prominent arc-like features trace the edge of the outermost
regions and are located at
PAs 64\degr--244\degr\ with brightness increasing with wavelength.

The color map of this source (Figure~\ref{Fig3}) shows values
  of temperature in the range from 150$\pm$5~K up to 390$\pm$5~K with a
mean value of 210~K. The maxima in temperature are related with regions at the tips of the major axis of
symmetry. The innermost ring-like structure surrounding the central star seems
to have two different temperature components: the south-west is the
coolest region ($\sim$150~K), whereas there is a hot component located toward
north-east with a temperature of $\sim$350~K. The 
arcs are warmer ($\sim$250~K). The mean value of the temperature
($T$=210~K) has been used to derive the optical depth maps of
IRAS\,17009$-$4154 (Figure~\ref{Fig4}).
The regions with the highest column densities are the arcs, followed by the ring-like structure, as shown 
in all optical depth maps.

\subsection{IRAS\,18229$-$1127}

IRAS\,18229$-$1127 has a rhomboidal clumpy envelope with an angular 
size of 2$\farcs$5. In the PAH1, SiC, and [Ne\,{\sc ii}] N-band
images (Figure~\ref{Fig2}), this envelope is dominated by two bright knots toward the 
north and north-west, and two weaker knots are noticeable toward the south
and south-east. The south-east knot is not detected in [Ne\,{\sc ii}]. Interestingly, these knots are singularly not apparent in the Q1 image that shows a
clearly hollow envelope, with a elongated cavity oriented along the north-south 
direction. The central star is detected,
but only in the bluest PAH1 image at 8.6 $\mu$m. The color-composite RGB picture of this object at
the bottom of Figure~\ref{Fig2} shows a reddened, hollow rhomboidal
envelope with a spot brighter at shorter
wavelengths at each corner. 

We have derived the color map of this source using its PAH1 and Q1
images and we estimate a mean temperature of 170~K (Figure~\ref{Fig3}), with a range of 
100~K$\leq T \leq$\,370$\pm$1~K. 

The bright knots in the N-band filters are the warmest regions of
this source, with $T\simeq$280~K, 
whereas the rest of the rhomboidal envelope has the lowest temperatures 
($\sim$200\,K). The mean value of  170~K has been used to derive the
optical depth maps of IRAS\,18229$-$1127 (Figure~\ref{Fig4}).
The knots represent the regions with the highest densities in the
N-band, with the north one being the most dense. These density variations are diminished in the Q1 optical depth map, 
where the rhomboidal envelope has a more homogeneous density
distribution. It is remarkable the difference between the optical depth maps of the
N-band and the one of Q1: while the knots are the densest zones from
8.6 to 12.8 $\mu$m, the
peak of the density at 17.7 $\mu$m is toward east, where not a
single knot is present.

\subsection{IRAS\,18454$+$0001}

This source has the smallest angular size among the objects in our
sample. The source can be described as a round, 1\farcs5 in size,
clumpy envelope in the SiC, [Ne\,{\sc ii}], and Q1 images, but
its morphology in the PAH1 image is singularly different
(Figure~\ref{Fig2}). In this
band, the image is dominated by a pair of bright knots along PA
121\degr\ that seem spatially coincident with brightness
peaks in the Q1 image, whereas the peaks detected in the SiC and
[Ne\,{\sc ii}] images are located at different positions. All images
hint at the presence of a central cavity, but its orientation and extent
varies with wavelength. At large spatial scale, the Q1
image reveals an extended halo around the main nebula.
The halo with a size of 5\arcsec$\times$4\arcsec\ extends toward the north-east, as
shown in red in the color-composite RGB picture at the bottom
Figure~\ref{Fig2}.

\begin{figure*}
   \centering
 \includegraphics[width=4.35cm,bb= 0 70 453 453]{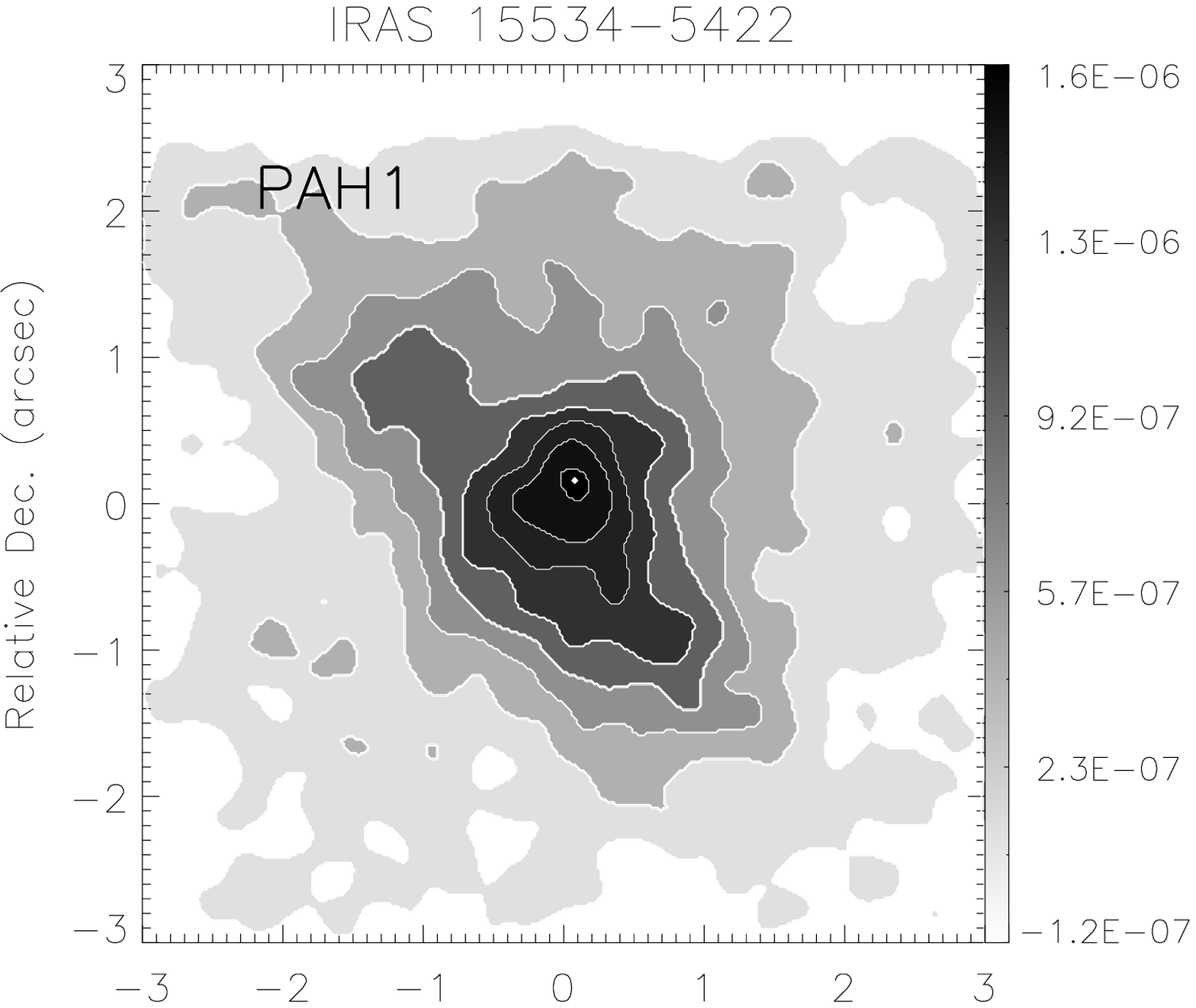}
   \includegraphics[width=4.35cm, bb= 0 70 453 453]{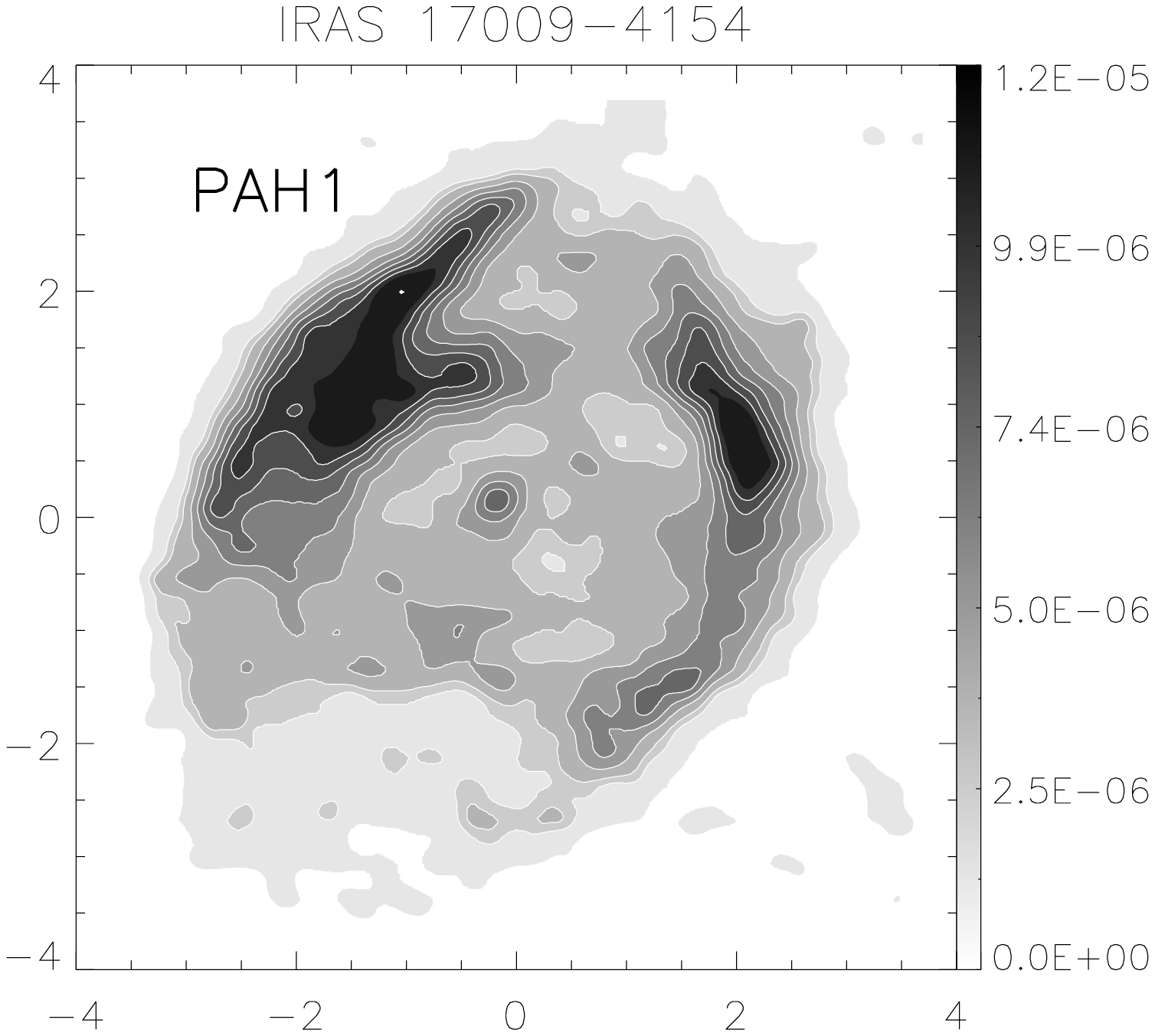}
   \includegraphics[width=4.35cm, bb= 0 70 453 453]{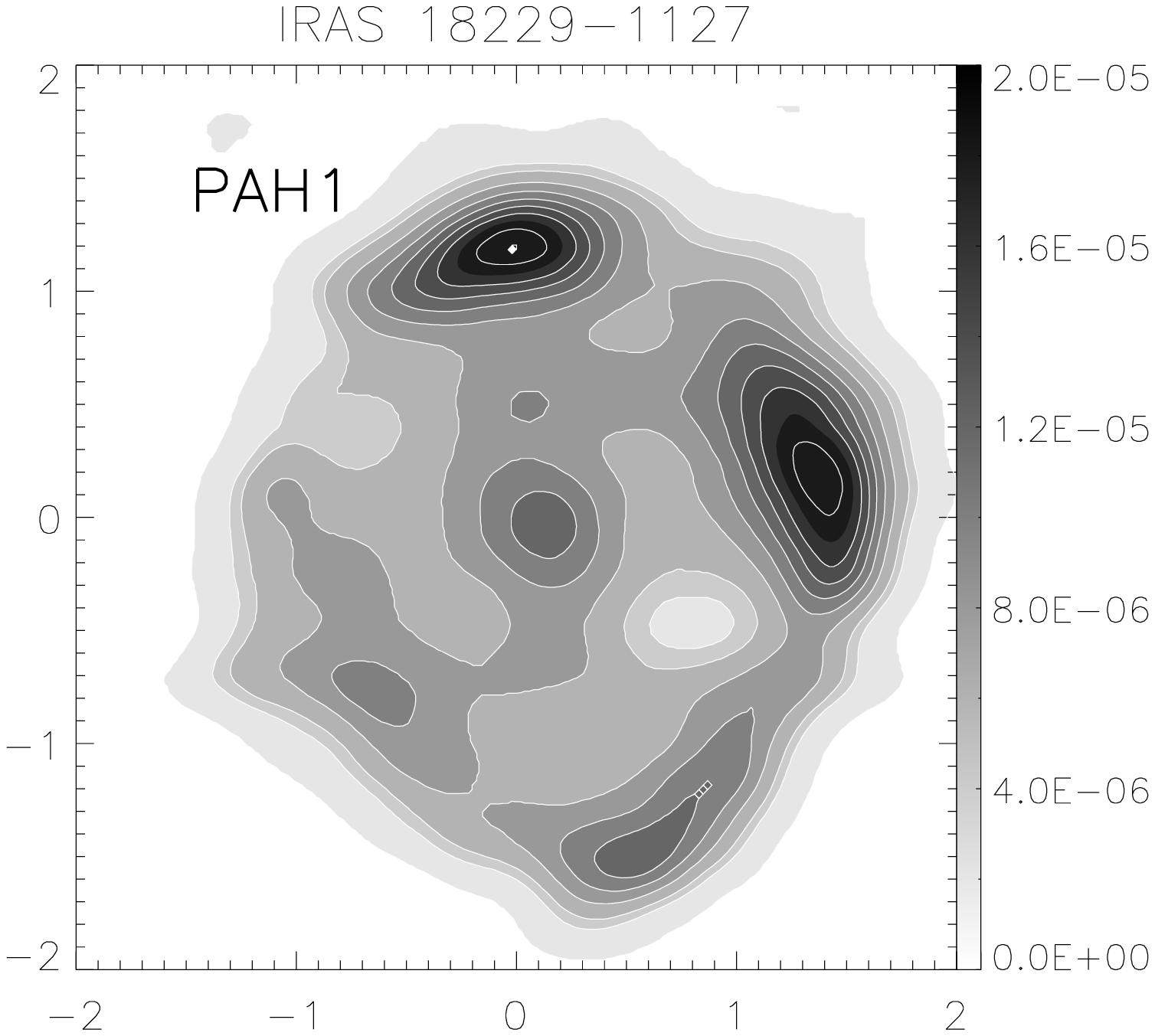}
   \includegraphics[width=4.35cm,bb= 0 70 453 453]{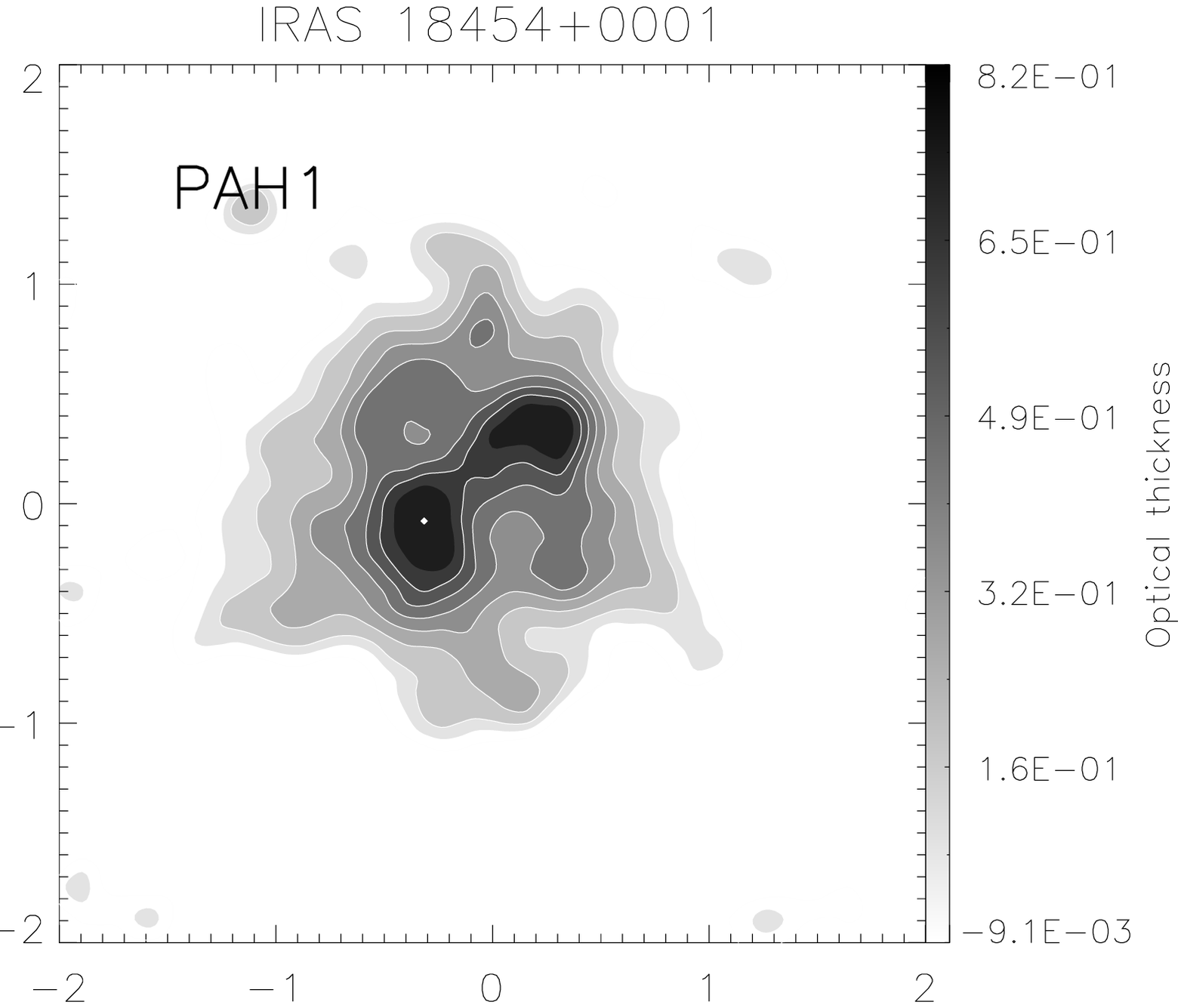}
   \includegraphics[width=4.35cm,bb= 0 70 453 453]{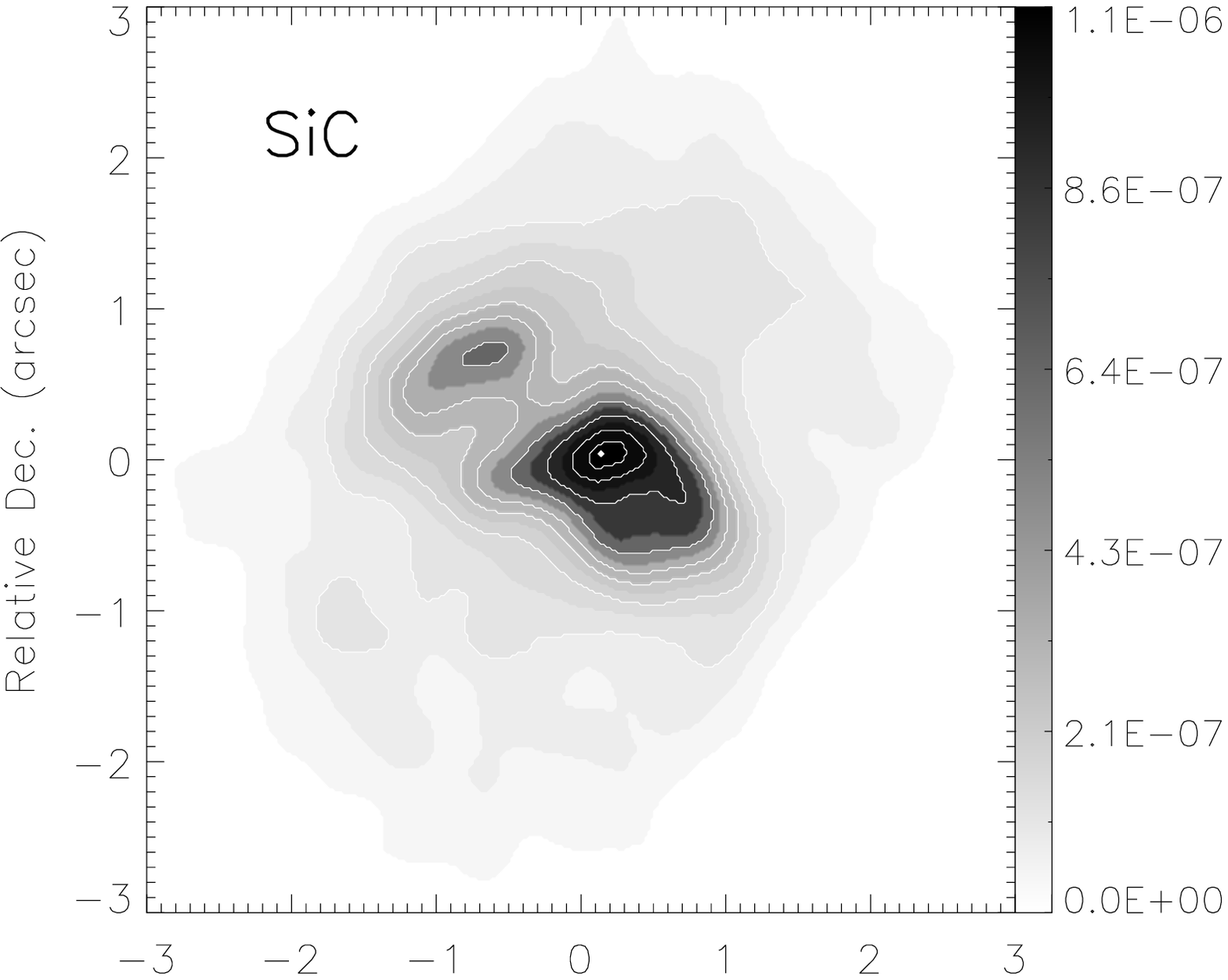}
   \includegraphics[width=4.35cm,bb= 0 70 453 453]{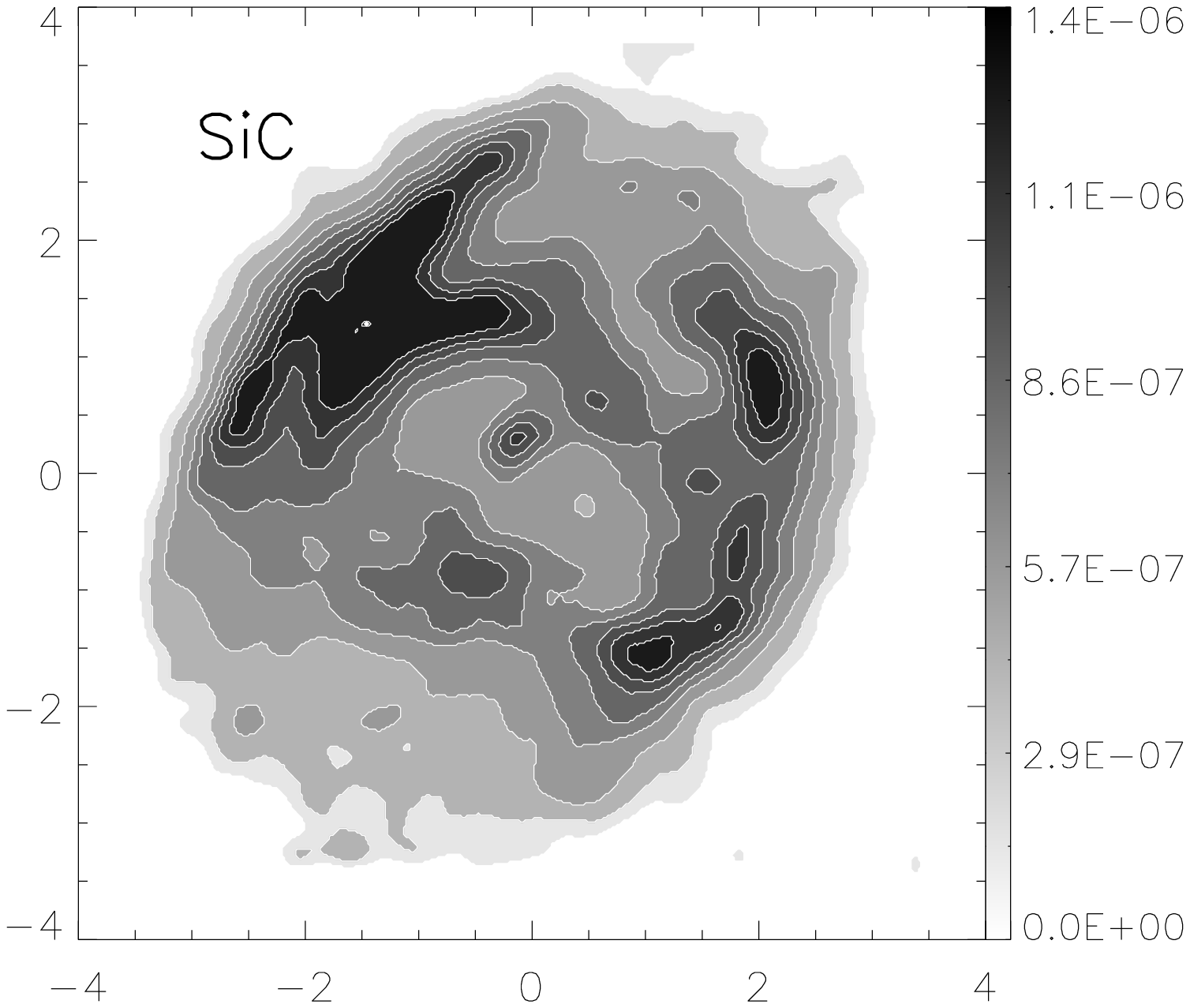}
   \includegraphics[width=4.35cm,bb= 0 70 453 453]{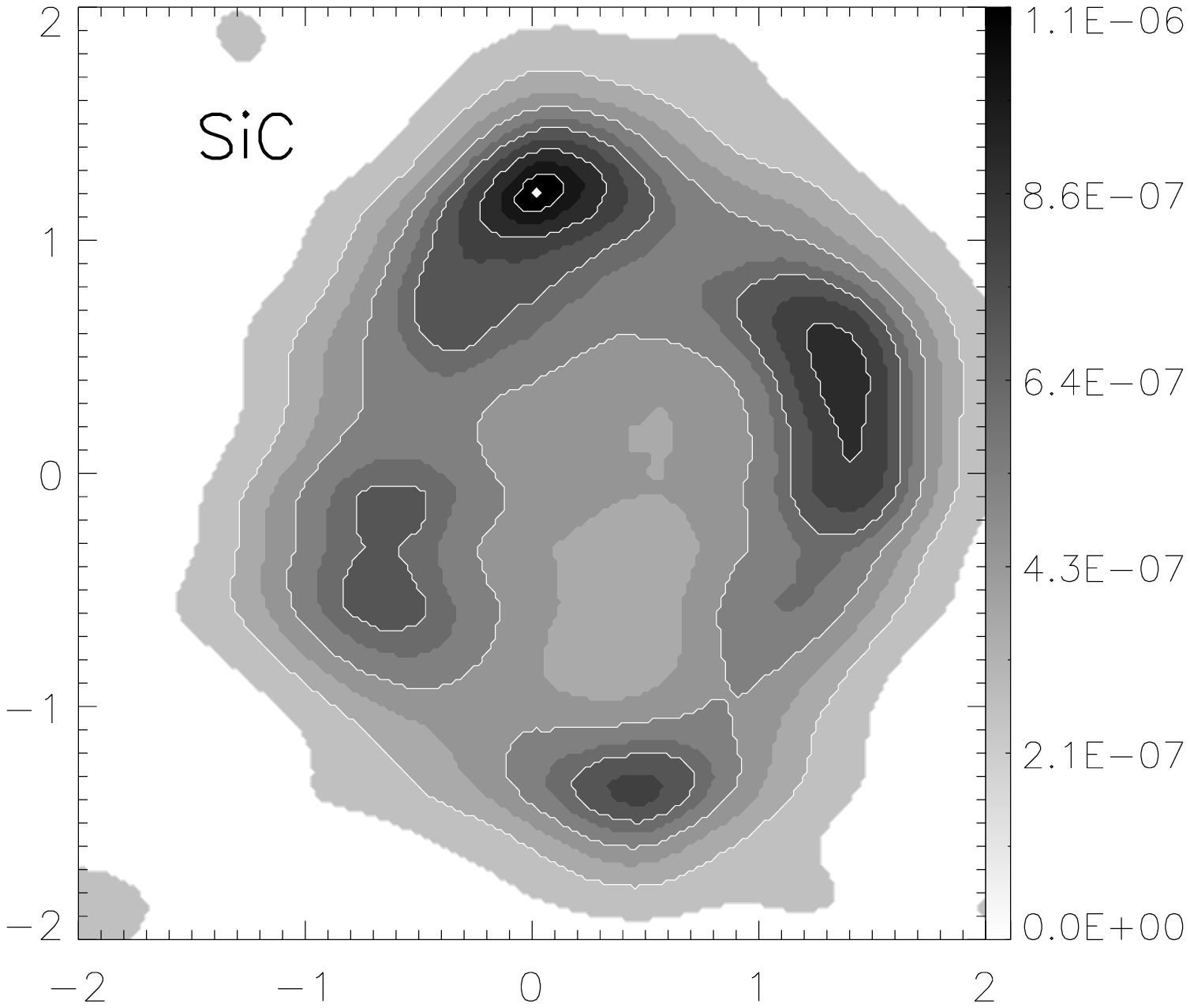}
   \includegraphics[width=4.35cm, bb= 0 70 453 453]{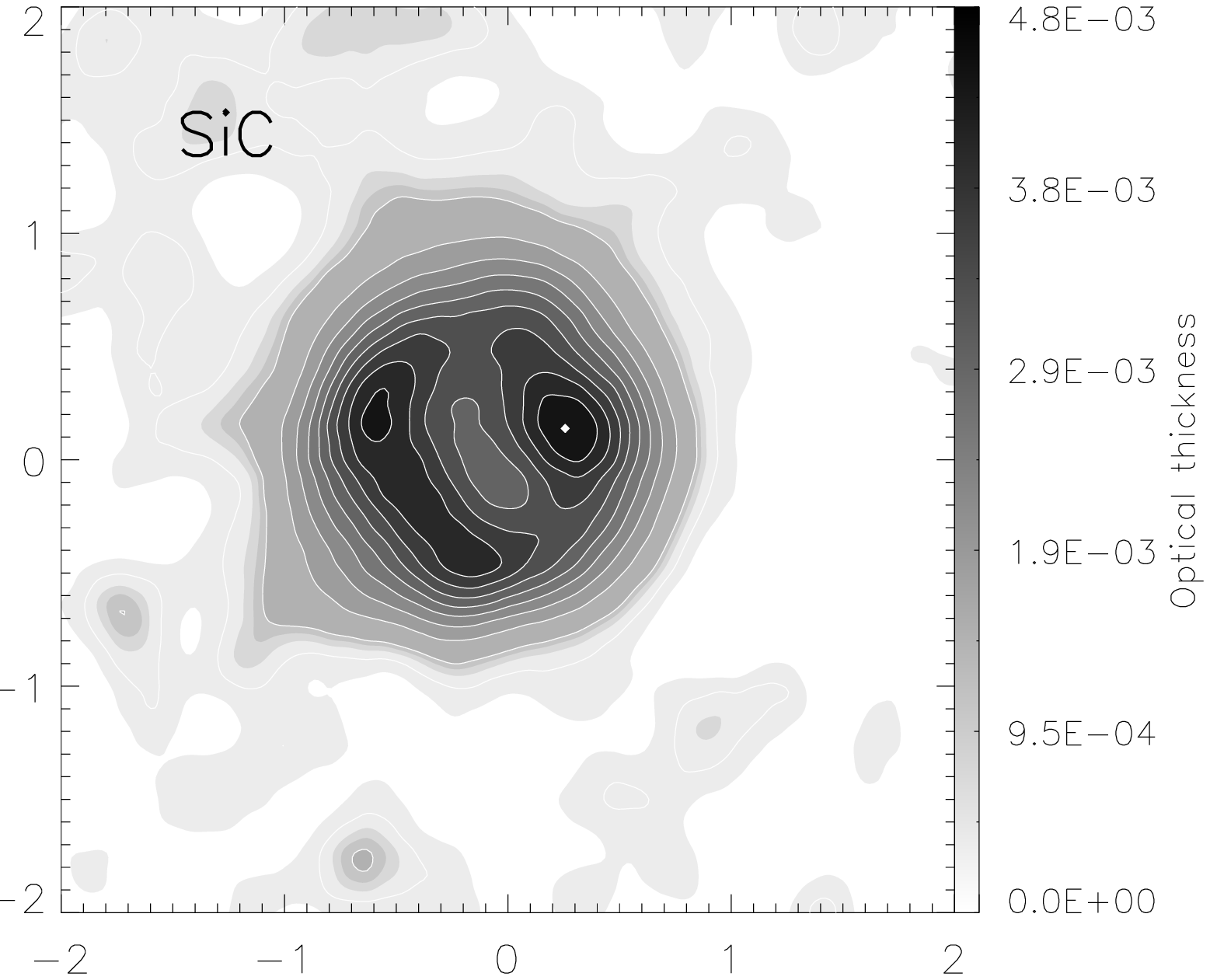}
   \includegraphics[width=4.35cm,bb= 0 70 453 453]{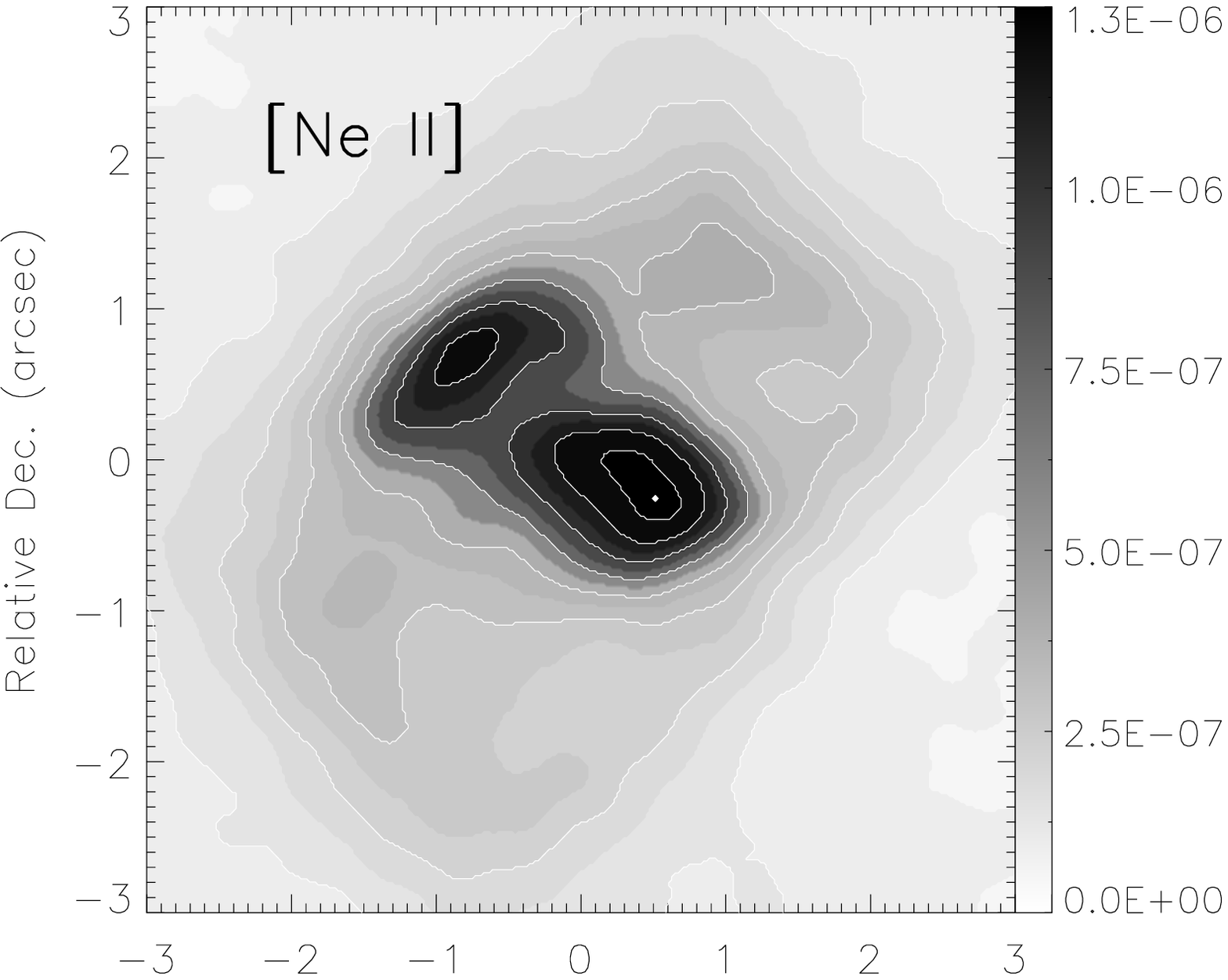}
   \includegraphics[width=4.35cm,bb= 0 70 453 453]{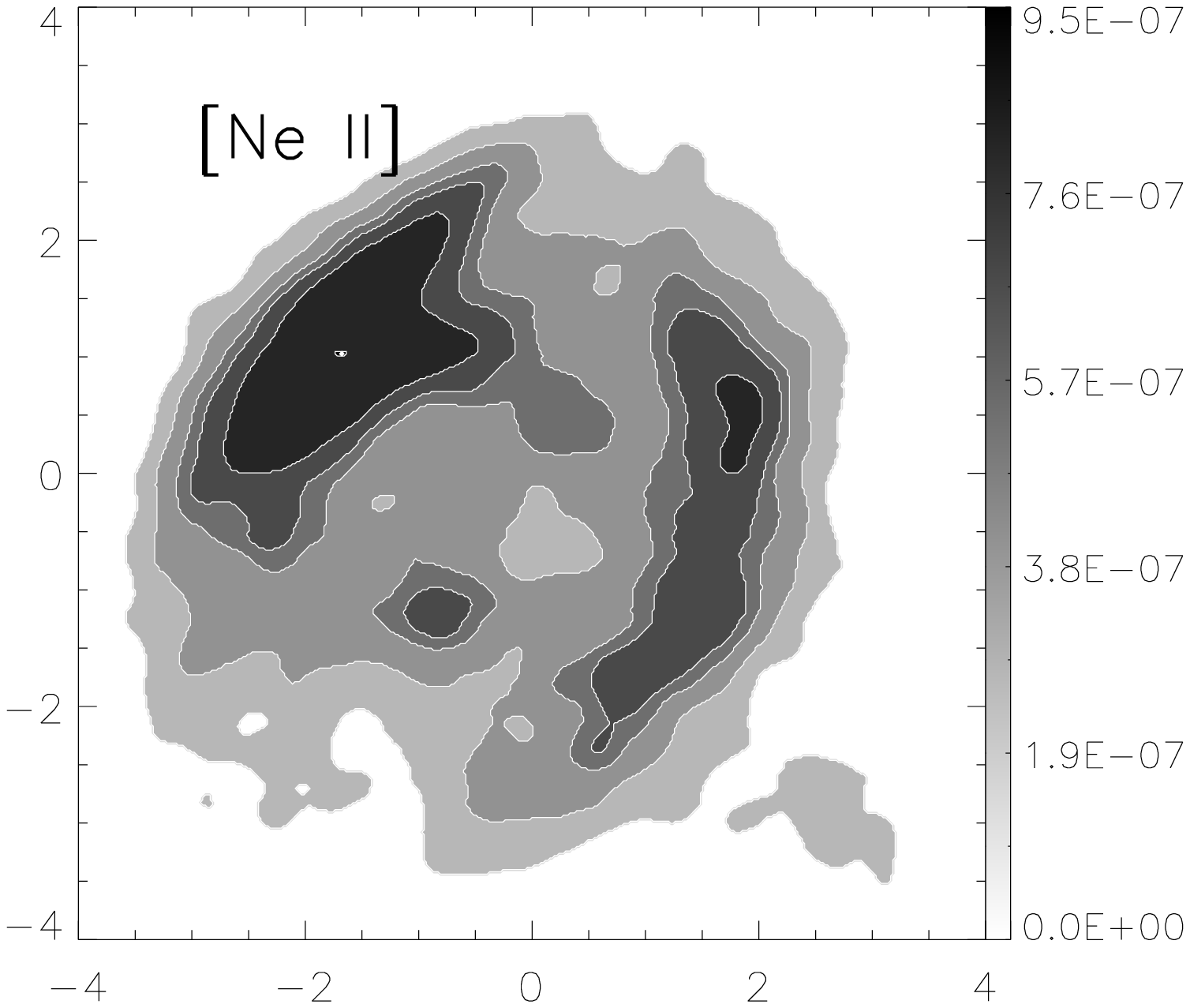}
   \includegraphics[width=4.35cm,bb= 0 70 453 453]{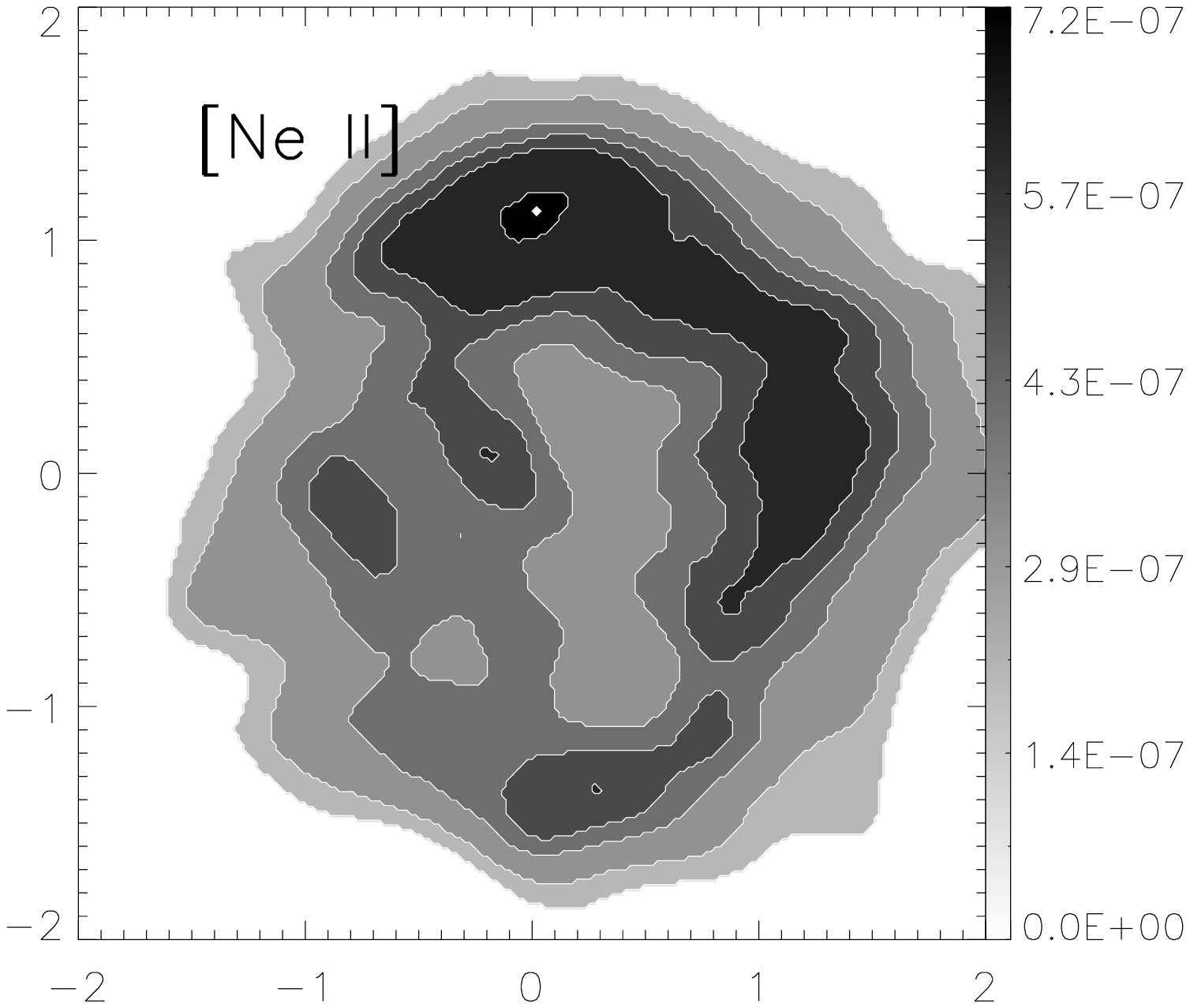}
   \includegraphics[width=4.35cm,bb= 0 70 453 453]{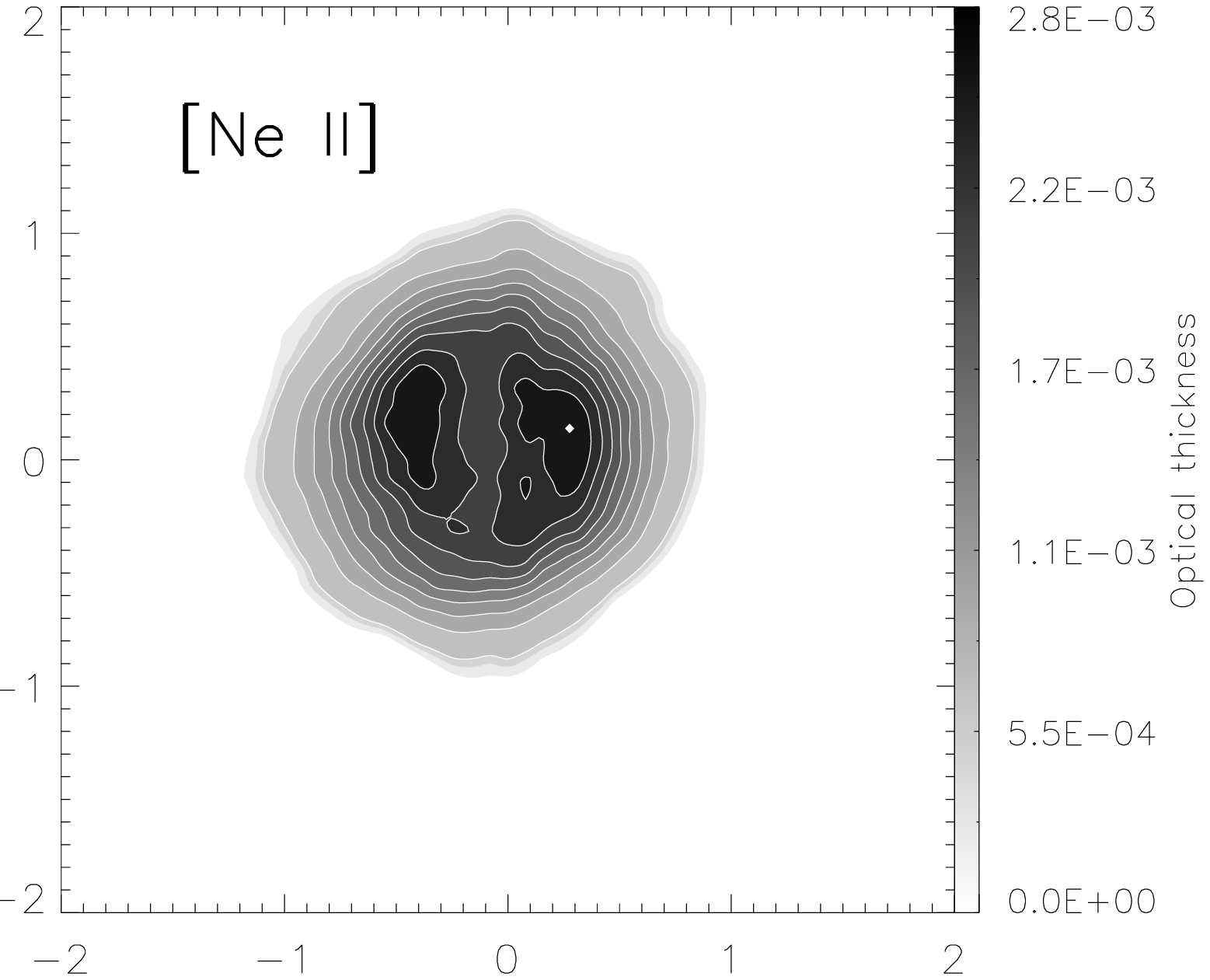}
   \includegraphics[width=4.45cm]{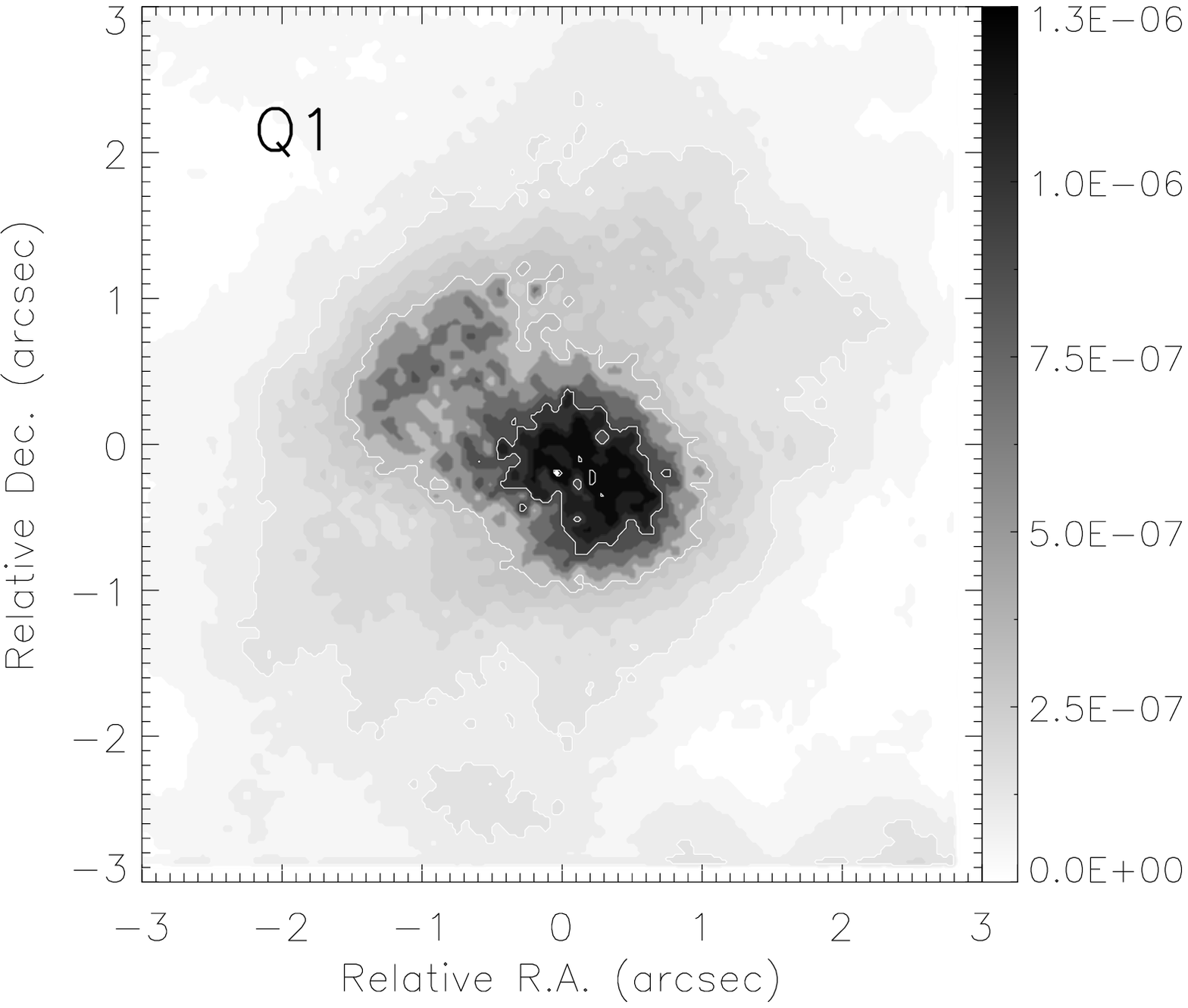}
   \includegraphics[width=4.35cm]{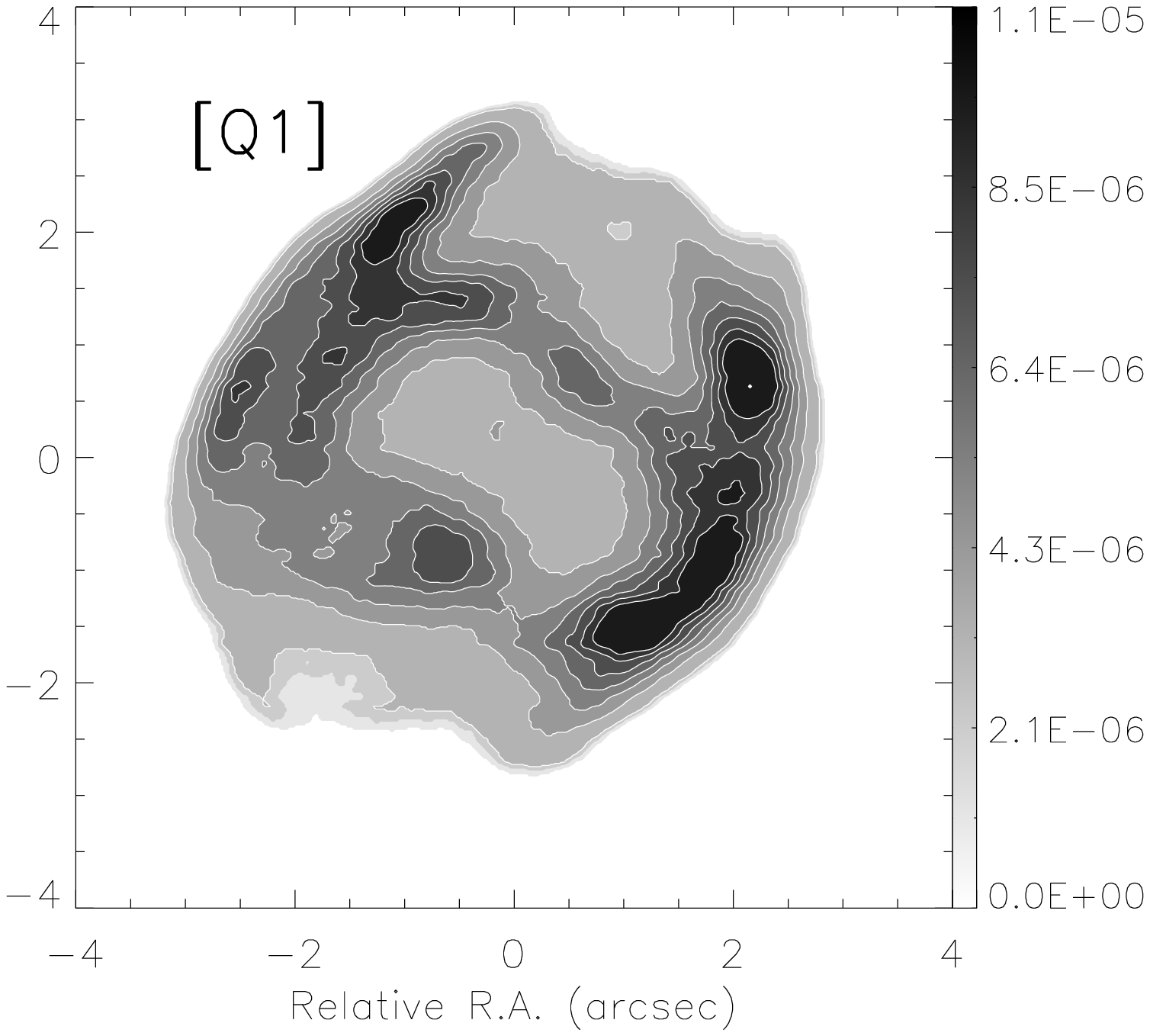}
   \includegraphics[width=4.35cm]{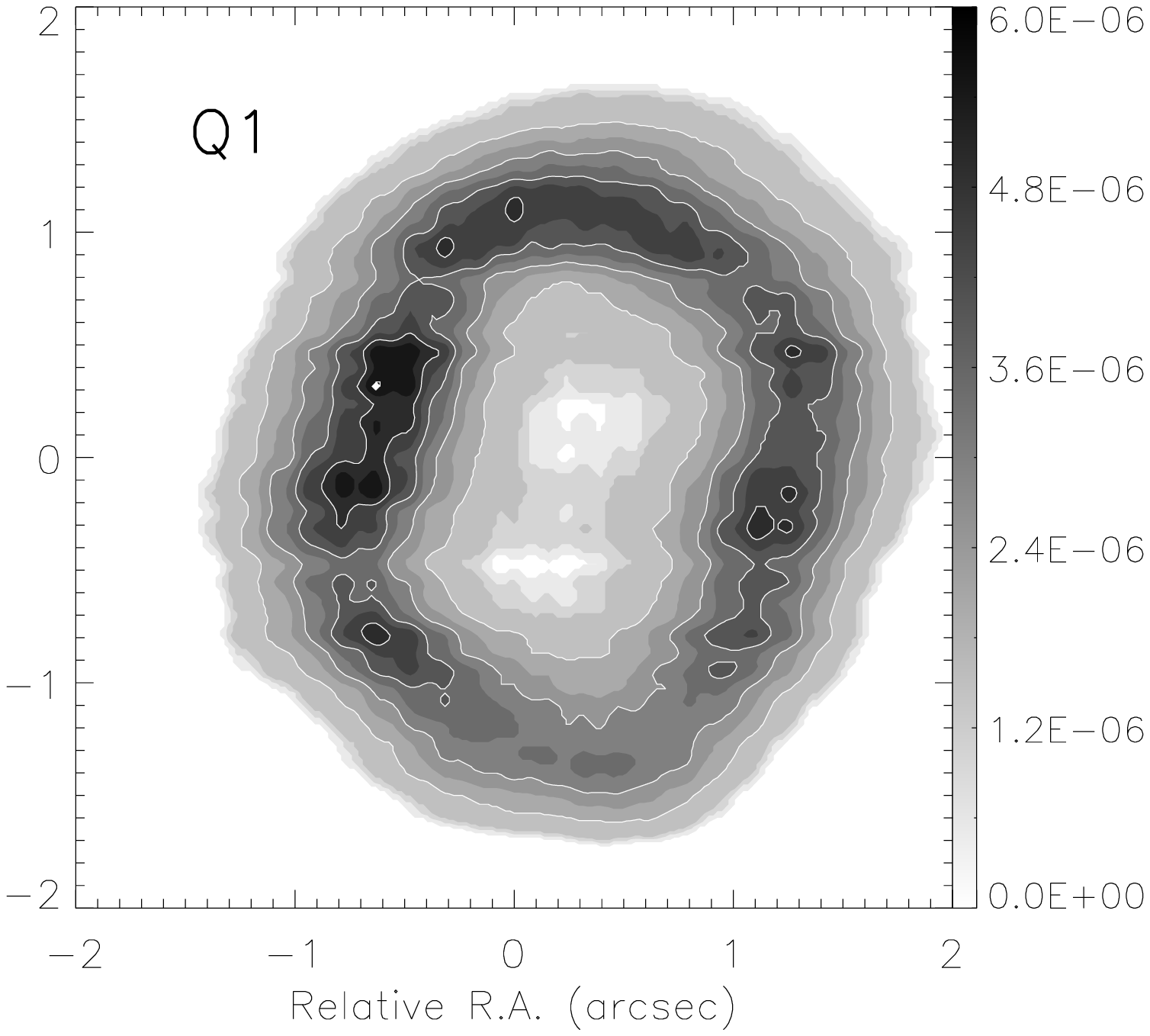}
   \includegraphics[width=4.35cm]{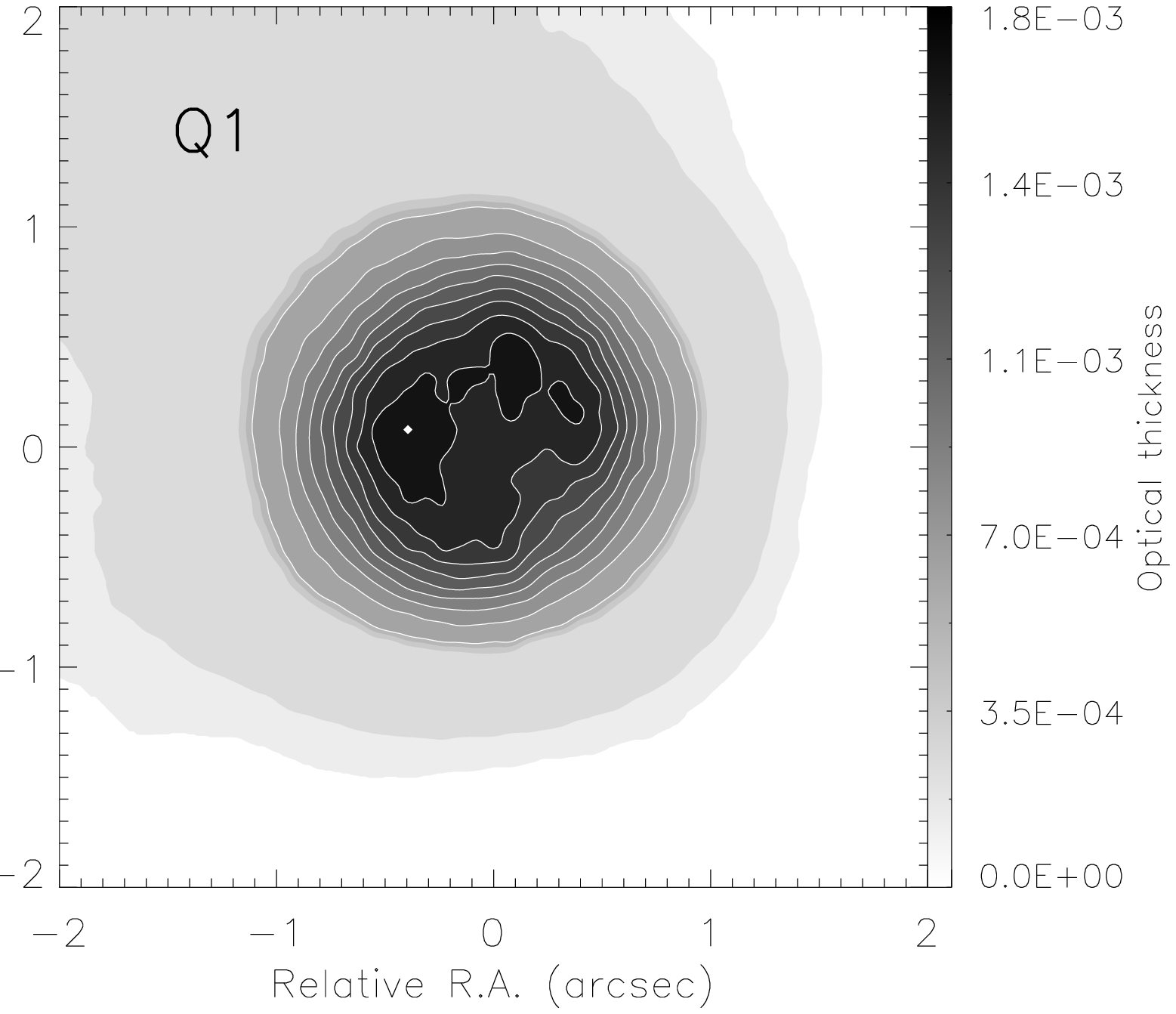}
 \caption{Optical depth maps of the sources obtained using
   equation 2 for the four different
   wavelengths observed. Contours are in
intervals of 10\% of the intensity peak for IRAS\,17009$-$4154,
IRAS\,18229$-$1127 and
IRAS\,18454$+$0001. For IRAS\,15534$-$5422 the contours are in
intervals at 10\%
for [Ne\,{\sc ii}] and SiC, intervals at 20\% for Q1, and for the PAH1
map the
contours start at 70\% of the intensity in intervals of
2.5\% of the intensity peak.
              }
         \label{Fig4}
   \end{figure*}

The [Ne\,{\sc ii}] and Q1 images of IRAS\,18454$+$0001 have been used
to estimate the mean value of temperature of 80~K in a range from
50$\pm$1~K to 120$\pm$1~K (Figure~\ref{Fig3}). 
The temperature variations are notably flat over the nebula, with the
hottest temperature ($\gtrsim$100~K) at the pair of bright spots seen
in [Ne\,{\sc ii}] and regions of high temperature ($\simeq$90~K) at
the bright PAH1 and Q1 knots. The emission of the envelope seems to be traced by a cooler dust 
component. Interestingly, the optical depth maps
(Figure~\ref{Fig4}) show that the 
highest density regions are associated with the knots seen in PAH1 
and in Q1, while the knots associated with the [Ne\,{\sc ii}] emission 
peaks represent zones of lower column density. There is a remarkable difference in the value and location of the
peaks in the the optical depth maps of IRAS\,18454$+$0001 in 
PAH1 and Q1 with respect to those in SiC and [Ne~{\sc ii}], but 
we note this may be an artifact produced by the low value of the 
temperature estimated for this source ($T$=80~K) that has been used to built 
its optical depth maps.

\section{Discussion}

\subsection{Physical structure of the sources}

We have obtained mid-IR high spatial resolution images of three
PNe and one post-AGB source candidate using VISIR at the VLT.
These images reveal notable morphological differences among the 
sources in our sample. 
The shell-like morphology and detection of a central star in three 
of these sources are not expected for the diffuse mid-IR emission 
from compact H~{\sc ii} regions \citep{Taka2000}.  
Only IRAS\,15534$-$5422 may look like a compact H~{\sc ii} region, but its 
PN nature is well established by optical spectroscopy \citep{Parker2012}.  

Temperature and optical depth maps built from
these images are proven to be powerful tools to enhance morphological
structures and to reveal new ones in these objects.  
In the following we will discuss the main characteristics of the 
individual sources as well as the characteristics that are shared by them 
all. 

IRAS\,15534$-$5422 is a young PN with two distinct structural 
components.  
The elongated innermost structure is dominated by thermal 
dust continuum and it represents the coldest and densest 
regions in this PN. 
This structure is surrounded by a warmer, ionized extended 
envelope of lower density. 
Whereas the color-composite picture of IRAS\,15534$-$5422 
in Figure~\ref{Fig2} seems to imply that the elongated inner structure is a 
bipolar nebula surrounded by the ring-like envelope, 
the physical properties of the different components favor an 
alternative, more solid interpretation.  
The innermost structure is rather a high-density, dusty ``torus'', 
whereas the apparent envelope can be associated to a pair of 
ionized bipolar lobes perpendicular to that ``torus''.  
This interpretation is consistent with the narrow-band near-IR 
images presented by \citet{RL2012} and with the near-IR excess 
noticeable in its SED (Fig.~\ref{Fig1}).  

IRAS\,17009$-$4154 seems to have an slightly bipolar or elliptical morphology.
The bright arcs can be interpreted as ionized bipolar extensions
of an elliptical shell sculpted by a SiC and Q1 bright, dusty
equatorial ring. 
The dust ring is the coldest structure in this source, whereas the bright 
arcs are warmer. The temperature is the highest at the 
northwest and southeast tips of the lobes.

Based on the properties of the SED of IRAS\,18229$-$1127 and on its
similarity with that of the PN IRAS\,18454$+$0001 (Fig.~\ref{Fig1}), 
we favor its classification as a post-AGB source and discard a
YSO nature.
This source shows a peculiar clumpy and dusty rhomboidal
envelope. This rhomboidal appearance is enhanced by the knots
located toward the northwest, the densest region of this source, and
southeast knots. The rhomboidal morphology of IRAS\,18229$-$1127
resembles the shape of the PN BD$+$303639 \citep{Lagadec2011,Akras2012}.

IRAS\,18454$+$0001 is a compact PN displaying both a spherical 
AGB shell and a dusty torus, most noticeable in the PAH1 (8.6~$\mu$m) and Q1
(17.7~$\mu$m) images. 
The temperature map suggests the presence of relatively hot regions
along the direction perpendicular to this torus. Alternatively, these 
regions may be dominated by ionized gas unveiled by its [Ne~{\sc ii}] emission.
The torus shows density enhancements in the dust-dominated PAH1 and 
Q1 filters corresponding with the high density knots along the
south-east 
and north-west direction.  
The presence of a high density torus, but a spherical AGB envelope, 
is intriguing.  
We might witness in this source the early phase on the shaping of a 
bipolar PN, before the spherical geometry of the AGB wind is disrupted 
by a fast stellar wind collimated by a toroidal density enhancement
as proposed by the Generalized Interacting Stellar Wind model of PN 
shaping \citep[GISW,][]{Balick87}.

The morphological features detected in direct images are notably
enhanced in the temperature and optical depth maps, as these
distinguish between structures of same brightness, but different 
physical conditions. Furthermore, the physical characteristics of the
structures detected for each source
are generally consistent. The innermost dusty rings or torii of IRAS\,15534$-$5422, IRAS\,17009$-$4154, and
IRAS\,18454$+$0001 have lower temperatures ($T\sim$100~K) than the
elongated hotter ($T\gtrsim$150~K) bipolar regions.

The optical depth maps of IRAS\,15534$-$5422, IRAS\,17009$-$4154, 
and IRAS\,18229$-$1127 peak at similar values for each of these 
sources.  
This implies that, there are no variations
in the density derived from the different filters for each of these sources, and thus that the 
mid-IR emission of these source is optically thin.

\subsection{Comparison with the morphology and physical structure of
  other evolved objects}

Our mid-IR VISIR-VLT images have unveiled a wealth of structural 
components in the four sources in our sample.
In short, these sources can be described as a detached rhomboidal
shell (IRAS\,18229$-$1127), a spherical shell with strong evidences 
of a dusty torus orthogonal to an ionized bipolar flow. 
(IRAS\,18454$+$0001), and two mild bipolar or elliptical sources 
(IRAS\,15534$-$5422 and IRAS\,17009$-$4154).
Contrary to other studies of sources in the transition towards the PN 
phase \citep{Sahai98,Ueta2000,Sahai07,Sahai2011,Lagadec2011}, we 
do not find evidence of extreme axisymmetric bipolar or multipolar
morphologies. 
In this respect, we note that we have not imposed to the sources in our 
sample any of the restrictions of previous studies, that required the 
sources to be detected in the optical \citep{Ueta2000,Sahai2011}, or to 
have fluxes at 12 $\mu$m greater than 10~Jy \citep{Lagadec2011}. 
These selection criteria may have introduced notorious biases
in the morphological output of the sample.

In this sense, the lack of restrictive selection criteria in our sample 
may probe a different population of sources in the transition between 
the late AGB and early post-AGB 
phases. The small number of objects in our sample does not allow us to
drawn firm conclusions, however, the observed morphologies may be
analyzed within the evolutionary context of PN formation. 
At least three of the sources in our sample (IRAS\,15534$-$5422, 
IRAS\,17009$-$4154, and IRAS\,18454$+$0001) have already reached 
the PN stage, although they are still mostly obscured at optical 
wavelengths.  
The significant obscuration can be attributed to large amounts of
circumstellar material that has been previously ejected by their central
stars, suggesting massive progenitors. Interestingly, the optical
depth values computed for IRAS\,18454$+$0001 are significantly larger
than those of the remaining sources, whereas its
temperature is lower. The thicker envelope of this source may be
indicative of an even more massive progenitor. 

If the sources in our sample indeed descended from massive AGB stars, 
we do not see the expected correlation between extreme bipolar morphology 
and massive progenitors \citep{Corradi95,Ueta2000,Siod2008}. 
It can be argued that the sources
in our sample are at a very early evolutionary phase, when asymmetries have not fully
developed yet. For example, IRAS\,15534$-$5422, with its dense,
dusty torus and bipolar lobes, might be a bipolar PN in the making, 
whereas the high density torus of IRAS\,18454$+$0001, enclosed 
within a spherical shell, can provide the seed for a bipolar PN. 

The use of color maps in the mid-IR domain has revealed a torus in
IRAS\,07134$+$1005 \citep{Dayal98} and hinted the presence of jets 
in Roberts~22 and in V~Hya \citep{Lagadec2005}. 
The mid-IR size of these sources ($\sim$4\arcsec) are similar to that 
of the sources in our sample, but the variations of the temperature 
across the sources studied in this work have shown structures that 
define the morphology more clearly. 
Compared with these previous analyses of color and optical depth maps, 
our study resolves much more morphological details, specially if we 
note that the sources in our sample do not display extreme
axisymmetric nebulosities.  
The high-spatial resolution achieved by our data is certainly required
to resolve the extended emission of the obscured and small-sized or 
compact sources transiting this evolutionary phase.

\section{Conclusions}

We have observed four heavily obscured post-AGB sources and PN candidates 
with VISIR-VLT in three different N bands (PAH1, SiC and [Ne~{\sc ii}]), 
and one Q band (Q1).  
Three sources in our sample (IRAS\,15534$-$5422, IRAS\,17009$-$4154, 
and IRAS\,18454$+$0001) can be classified as young PNe, whereas only 
a preliminary classification as a post-AGB source is possible for IRAS\,18229$-$1127.

The high-spatial resolution VISIR images have been used to investigate the 
extended emission and to study the spatial variations of the physical 
conditions (temperature and optical depth) of these sources.  
We are providing evidence of asymmetry in three young PNe: 
an innermost dust torus or ring embedded within an ionized 
spherical shell (IRAS\,18454$+$0001), and two mild bipolar 
or elliptical sources with dusty rings (IRAS\,15534$-$5422 and 
IRAS\,17009$-$4154).  

Compared to previous works, the use of color and optical depth maps 
have proven much useful to reveal fine structural details in a small 
sample of heavily obscured sources reaching the PN phase, 
confirming the usefulness of mid-IR high resolution observations 
for the study of this short evolutionary phase.

Our imaging study confirms that asymmetry is present in heavily obscured 
post-AGB stars and young PNe but, contrary to previous studies
focused in these evolutionary phases, no extreme axisymmetric morphologies are found.  
These previous studies may be biased towards the mid-IR and optical 
brightest sources, missing critical early phases of the post-AGB 
evolution of the most massive progenitors.  
Our study, although based on a small sample, may be yielding important 
clues on the onset of asymmetry in massive progenitors of PNe.
Further studies of the most obscured post-AGB sources must be 
pursued in the future.

\begin{acknowledgements}
      Part of this work was supported by the Ministerio de Econom\'ia
      y Competitividad of Spain through grants AYA~2008-01934, 
AYA~2008-06189-C03-01, AYA~2011-29754-C03-02, and 
AYA~2011-30228-CO3-01 cofunded by FEDER funds. 
MWB would like to thank to the EEBB-FPI for the grant to perform the short term
stay in 2011, and to the European Southern Observatory (ESO)
Headquarters in Garching, Germany, for all the facilities provided
during this stay. 
GRL acknowledges support from CONACyT (grant 177864) and PROMEP 
(Mexico). 
LFM is also supported by grant IN845B-2010/061 of Xunta de
Galicia, partially funded by FEDER funds. 
This paper made use of information from the red \emph{MSX} Source survey database at
www.ast.leeds.ac.uk/RMS which was constructed with support from the
Science and Technology Facilities Council of the UK. We would like to thank the referee, Dr.\ C.\ Waelkens for his useful comments
for the improvement of this paper.
\end{acknowledgements}

\bibliographystyle{aa} 

\end{document}